\documentclass[lettersize,journal]{IEEEtran}
\usepackage{float}
\usepackage[caption=false,font=normalsize,labelfont=sf,textfont=sf]{subfig}
\pdfoutput=1
 
\usepackage{cite}
\usepackage[pdftex]{graphicx}
\usepackage[section]{placeins}
\usepackage{tabularx}
\usepackage{siunitx}
\usepackage{amsmath, amssymb,mathrsfs, dsfont} 
\usepackage{amsmath,amssymb,bm,bbm}
\usepackage{graphics,graphicx}
\usepackage{algorithm} 
\usepackage{algpseudocode} 
\usepackage{multirow}
\usepackage{hyperref}
\usepackage{soul}

\usepackage{xfrac}  
\usepackage{amsmath}
\usepackage{subfiles} 
\usepackage{booktabs}
\usepackage{amsfonts}
\usepackage{booktabs}
\usepackage{siunitx}
\usepackage[dvipsnames]{xcolor}
\usepackage[deletedmarkup=sout,authormarkup=superscript]{changes}

\usepackage{caption}
\definechangesauthor[name={Mahdi}, color=NavyBlue]{MN}
\definechangesauthor[name={Mohammad}, color=RedViolet]{MK}
\definechangesauthor[name={Alisa}, color=Orange]{AR}
\definechangesauthor[name={John}, color=Brown]{JL}

\graphicspath{{\subfix{../figures/}}}

\usepackage{lineno}
\begin{document}
\title{Guided Bayesian Optimization:\\ Data-Efficient Controller Tuning with Digital Twin}

        
\author{Mahdi Nobar,
        J\"urg Keller, 
        Alisa Rupenyan, 
        Mohammad Khosravi,
        and John Lygeros
\thanks{The Swiss National Science Foundation supported this work through NCCR Automation, grant number 180545. (Corresponding author: Mahdi Nobar.)}
\thanks{Mahdi Nobar and John Lygeros are with the Automatic Control Laboratory, ETH Z\"urich, Zurich, Switzerland (email: mnobar@ethz.ch; jlygeros@ethz.ch).}
\thanks{J\"urg Keller is with the Automation Institute, FHNW, Windisch, Switzerland (email: juerg.keller1@fhnw.ch).}
\thanks{Alisa Rupenyan is with the ZHAW Centre for AI, Zurich University for Applied Sciences, Zurich/Winterthur, Switzerland (email: alisa.rupenyan@zhaw.ch).}
\thanks{Mohammad Khosravi is with the Delft Center for Systems and Control, Delft University of Technology, Delft, Netherlands (email: mohammad.khosravi@tudelft.nl).}}

\markboth{IEEE TRANSACTIONS ON AUTOMATION SCIENCE AND ENGINEERING}
{Mahdi Nobar, \MakeLowercase{\textit{(et al.)}: Guided Bayesian Optimization: Data-Efficient Controller Tuning with Digital Twin}}

\maketitle

\begin{abstract}
This article presents the guided Bayesian optimization (BO) algorithm as an efficient data-driven method for iteratively tuning closed-loop controller parameters using a digital twin of the system. The digital twin is built using closed-loop data acquired during standard BO iterations, and activated when the uncertainty in the Gaussian Process model of the optimization objective on the real system is high.
We define a controller tuning framework independent of the controller or the plant structure.
Our proposed methodology is model-free, making it suitable for nonlinear and unmodelled plants with measurement noise.
The objective function consists of performance metrics modeled by Gaussian processes.
We utilize the available information in the closed-loop system to progressively maintain a digital twin that guides the optimizer, improving the data efficiency of our method.
Switching the digital twin on and off is triggered by our data-driven criteria related to the digital twin's uncertainty estimations in the BO tuning framework.
Effectively, it replaces much of the exploration of the real system with exploration performed on the digital twin.
We analyze the properties of our method in simulation and demonstrate its performance on two real closed-loop systems with different plant and controller structures.
The experimental results show that our method requires fewer experiments on the physical plant than Bayesian optimization to find the optimal controller parameters. 
\end{abstract}
\def\abstractname{Note to Practitioners}
\begin{abstract}
Industrial applications typically are difficult to model due to disturbances.
Bayesian optimization is a data-efficient iterative tuning method for a black box system in which the performance can only be measured given the control parameters.
Iterative measurements involve operational costs.
We propose a guided Bayesian optimization method that uses all information flow in a system to define a simplified digital twin of the system using out-of-the-box methods.
It is continuously updated with data from the system. 
We use the digital twin instead of the real system to perform experiments and to find optimal controller parameters while we monitor the uncertainty of the resulting predictions.
When the uncertainty exceeds a tolerance threshold, the real system is measured, and the digital twin is updated.
This results in fewer experiments on the real system only when needed.
We then demonstrate the improved data efficiency of the guided Bayesian optimization on real-time linear and rotary motor hardware. These common industrial plants need to be controlled rigorously in a closed-loop system.
Our method requires $57\%$ and $46\%$ fewer experiments on the hardware than Bayesian optimization to tune the control parameters of the linear and rotary motor systems.
Our generic approach is not limited to the controller parameters but also can optimize the parameters of a manufacturing process.
\end{abstract}

\begin{IEEEkeywords}
Learning control systems, Iterative methods, Database systems, Control systems, Optimization methods
\end{IEEEkeywords}
\IEEEpeerreviewmaketitle
\vspace{-5 pt}
\section{Introduction}\label{sec:introduction}

\IEEEPARstart{M}{echatronic} systems require periodic retuning to deal with the uncertainties caused by the system's operation.
The controller parameters are typically conservative in handling the unknown and time-varying characteristics of the industrial systems.
Auto-tuning methods for adjusting the controller parameters replace the awkward manual controller tuning \cite{borase2021review, TASE_MPC_Tuning}.
Classical model-based controllers derive a model for the system by identification or first-principle methods.
The symmetric optimum tuning methods use a performance metric such as phase margin or the closed-loop system's bandwidth to tune the controller \cite{astromfeedback,TASE_high_bandwidth_tuning}.
Global optimization algorithms typically require a precise plant model or many trials.
For example, given a performance-based objective function in a fixed boundary, particle swarm optimization \cite{kennedy1995particle} determines appropriate controller gains\cite{PS_8799023}.
The genetic algorithm used to tune PID controller parameters dealing with complex cost functions where the gradient-based methods struggle \cite{GA2_MAGKOUTAS2023104788, GA_Blondin2021}.  

One can directly incorporate available plant data in the control parameter tuning and the control algorithm which can relieve the modeling task of complicated systems.
Convolution-based data-driven method \cite{10297976} simulates the feedback system without building any model of it that accounts for the controller nonlinearities.
This method can control rotary actuators by accurately estimating the control input and system response\cite{10505715}, but it holds a strong assumption that the plant is a linear time-invariant system.
Reachable zone-based pure data-driven controller design uses only noisy input-output data of a trajectory of the system \cite{alanwar2021robust} independent from the system's model for controller tuning.
Nevertheless, such direct methods are usually less data-efficient than their model-based counterparts, especially when the controller needs to learn about complex real systems with limited data \cite{Dulac-Arnold2019}.

Integrating the measurements with simulation behaviors allows the data-driven controller to adapt effectively to the real world \cite{Schoettler2020}.
Learning from measurement can improve the tracking accuracy of an inverse compensator for a fiber-based piezoelectric actuator \cite{CHEN2024118088}.
Furthermore, the iterative learning controller combines the process measurements with the optimization framework to provide robustness in the presence of repetitive disturbances and plant-model mismatch \cite{afkhami2023robust, wang2023precise, DominicLM}.
A digital twin (DT) is a virtual replica of a real-world system that continuously updates with measurements from its physical counterpart.
DT enables online monitoring and optimization of complex manufacturing processes.
For example, \cite{jiang2021industrial} explains how a liquid-level control system can be monitored by a DT that encompasses virtual twins of all subsystems.
These components are then synchronized with the physical counterpart through a proper bridge between the subsystems such that one can extend the application to control the real process variables \cite{9300250, LUO2020101974}.
However, there is a gap between the "twinning" property of DTs in industrial systems which limits the DT accuracy.
To mitigate the system-twin gap in the controller tuning, \cite{10189222} incorporates a generic learning algorithm into the DT that enables active braking control of a race car. 
\cite{10003111} challenges the sample efficiency of the learning-based algorithms in comparison with the local gradient-based black-box optimization methods.
\cite{TASE_RL_model_safety} uses the model-based policy iteration to ensure regional optimality by locally approximating the system's behavior.

Industrial closed-loop systems are required to satisfy high precision and performance standards.
The controllers in such systems typically contain a set of parameters.
One can also measure the performance as a cost function containing pre-defined metrics based on the system response to a specific signal.
Model predictive control (MPC) relies on a model of a system to tune controller parameters that optimize a cost function.
For instance, to train MPC parameters on a complex system, \cite{TASE_datadrivenMPC} defines multiple models to predict a nonlinear system's output using several sampling times.
However, a reliable plant model rarely exists, and identifying a potentially nonlinear industrial system in a closed loop is often not feasible without adequate system excitation.
To control a hard-to-model industrial process, \cite{TASE_kalmanfilteradaptive} uses the process input and output data to train a controller with a Kalman filter, reducing the measurement noise effect.
\cite{rupenyan2021performance} tunes such control parameters purely by data acquired while the system operates with repetitive tasks.
However, the high operation cost of such industrial systems demands data-efficient controller tuning approaches.

Bayesian optimization (BO) is a data-driven technique that iteratively tunes the controller parameters in a limited number of experiments using a \textit{surrogate model} of the cost function \cite{garnett_bayesoptbook_2023}.
In complex industrial applications, conducting each of the iterative experiments can involve time-consuming measurements \cite{dey2019systematic}.
To increase the BO controller tuning efficiency, one can integrate all available process information into the optimizer \cite{khosravi2019HP}.
For instance, \cite{guidetti2021plasma} uses coarse physical knowledge about the Plasma Spray Process to encapsulate a linear dependence between the process inputs in the prior mean distribution.
In \cite{9896930} the process information is incorporated in BO to save experimental time in a tedious additive manufacturing application.
Furthermore, \cite{TASE_goal_oriented_MPC_BO} uses BO to tune the parameters of a goal-oriented model predictive controller that sets the reference input to another cascaded control system to reduce the energy consumption of a thermal management system.
This way, the predictive controller's state-space model serves as prior knowledge for BO, ruling out the search space that contradicts the plant's physics.
In previous examples, BO is guided by prior information about the process.


This work proposes \textit{Guided Bayesian Optimization} (Guided BO) method that incorporates available closed-loop system information into the optimization.
Our data-efficient method reduces the required real system operation to tune the industrial feedback controller parameters iteratively.
In our approach, automated controller tuning seeks a global optimum of a cost function that estimates the desired closed-loop system behavior.
We use the available data without additional system operation to create a DT.
We only take measurements on the real system when the BO surrogate model has sufficiently low uncertainty.
Otherwise, we use our DT to estimate the system performance and guide the optimizer.
This way, we tune the controller parameters with fewer experiments on the real system, reducing the operating cost.
We demonstrate the enhanced performance of the guided BO algorithm in simulation and two real industrial feedback control systems.

Section \ref{sec:Controller_Tuning_Problem} introduces the data-driven model-free controller tuning problem.
Then, in Section \ref{sec:Guided_BO}, we review the BO method and present our guided BO algorithm.
Section \ref{sec:System_Modelling} provides preliminary information about the system and controller structure, the performance metrics, and the overall cost function used for the numerical studies in \ref{sec:Numerical_Analysis}.
Finally, we demonstrate our guided BO method on two experimental setups in \ref{sec:Exper_Results}.


\vspace{-5 pt}
\section{Data-Driven Controller Tuning Problem}\label{sec:Controller_Tuning_Problem}
\begin{figure}[!t] 
\centering
\includegraphics[width=3.5in, trim=2.2cm 3cm 2.2cm 3cm, clip]{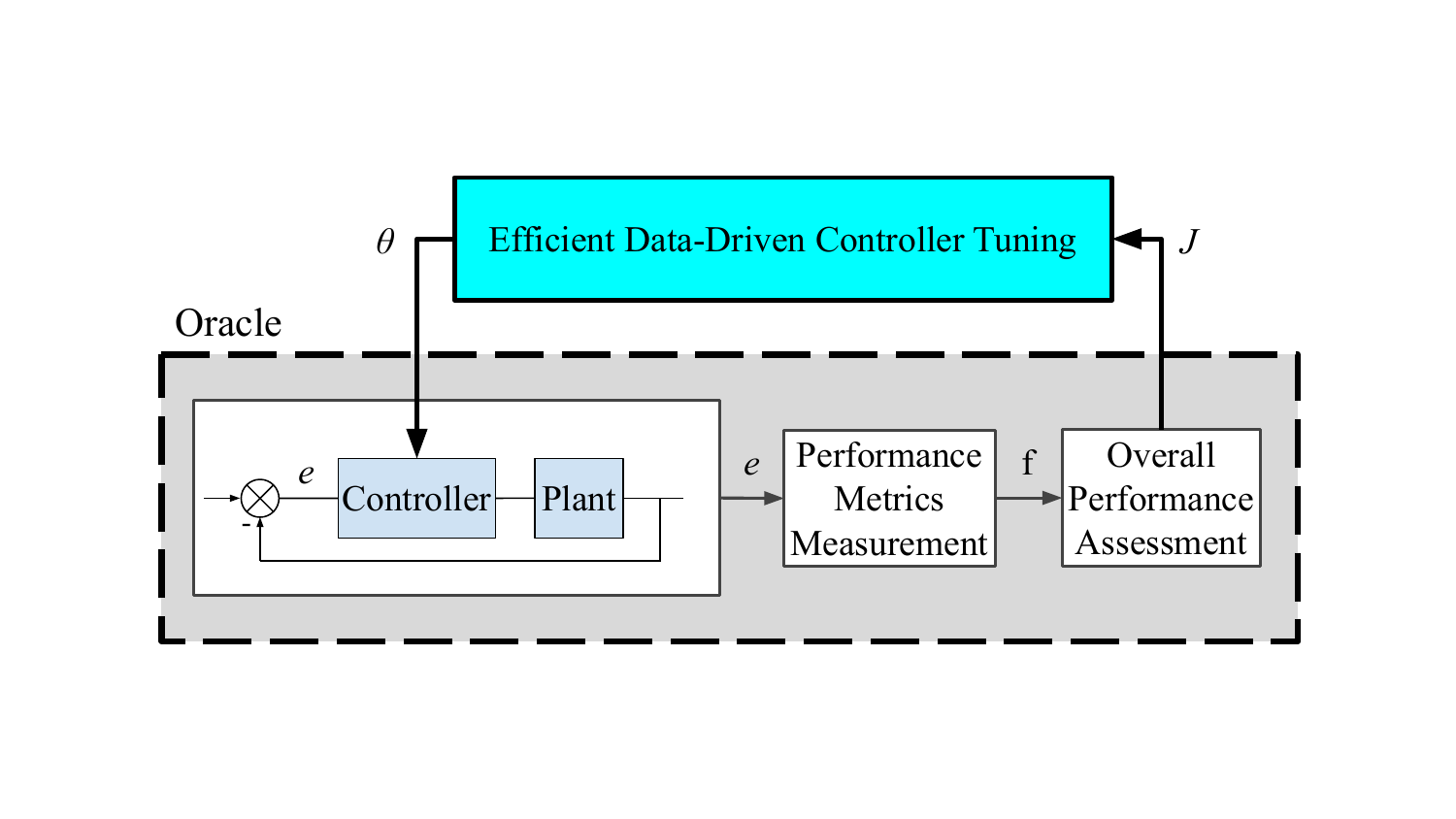}
\caption{Efficient data-driven controller tuning scheme based on performance assessment}
\label{fig:controller_tuning_problem}
\end{figure}
We consider a feedback control system with possibly a nonlinear plant.
Fig. \ref{fig:controller_tuning_problem} visualizes the components of our controller tuning problem. 
We have no assumption on the parametric controller structure.
Let $\theta \in \Theta$ be the \textit{controller parameter vector} inside a feasible set $\Theta \subseteq \mathbb{R}^{n_{\theta}}$.
We assume the feasible set $\Theta$ is known (see \ref{subsec:nominal_tuning} and \ref{subsec:linear_servo_motor_system}).
For the convergence properties of the closed-loop system, we choose the set $\Theta$ such that the step response of the system is bounded in a fixed period.

Given the controller parameter vector $\theta$, we denote the cost function $\hat{J}(\theta)$, where $\hat{J}:\Theta\to\mathbb{R}$ is a scalar metric reflecting the closed-loop performance.
There is no assumption on the convexity of the cost function.
So $\hat{J}(\theta)$ may have multiple local minima.
We want to simultaneously optimize multiple performance metrics denoted by $f_{i}: \Theta \rightarrow \mathbb{R}$ for $i=1,...,n_{\mathrm{f}}$ and $n_{\mathrm{f}}>1$.
These metrics are measured based on the tracking error signal $e$, the difference between the reference and output signal.
While one approach would be to optimize all the metrics separately, determining the Pareto frontier\cite{MObjBO}, we adopt the classical weighted aggregation method to convert our multi-objective problem to a single-objective problem \cite{ParetoFront}.
So the overall performance cost function $\hat{J}(\theta)$ is defined as 
\begin{equation}\label{eq:J}
\hat{J}(\theta):=\mathrm{w}^\mathsf{T}\mathrm{f}(\theta)=\sum_{i=1}^{n_{\mathrm{f}}}w_{i}f_{i}(\theta),
\end{equation}
where $\mathrm{f}(\theta):=[f_{1}(\theta),...,f_{n_{\mathrm{f}}}(\theta)]$ and $\mathrm{w}:=[w_{1},...,w_{n_{\mathrm{f}}}]$.
Since the data collecting process contains uncertainty, we assume that $J(\theta):=\hat{J}(\theta)+ \epsilon$ where $J(\theta)$ is the value of $\hat{J}(\theta)$ computed from the noisy data and $\epsilon \sim \ \mathcal{N}(0, \sigma_{\epsilon}^{2})$ is the measurement noise realization drawn from a normal distribution with zero mean and variance of $\sigma_{\epsilon}^{2}$.
The weights vector $\mathrm{w}$ adjusts our metrics' priority and scale such that each metric properly contributes to the overall performance.

For practical reasons, we assume that system identification is not feasible to obtain a realistic plant model.
For instance, data acquisition from an industrial plant is typically expensive due to operational time and costs.
Consequently, we cannot calculate the cost function $\hat{J}(\theta)$ analytically. 
So we suppose a \textit{black-box oracle} for our closed loop system that provides the cost value $\hat{J}$ given the controller parameter vector $\theta$.
Now our \textit{data-driven controller tuning problem} is to obtain the optimum controller parameter vector $\theta^{*}$ by solving
\begin{equation}\label{eq:opt_prob}
\begin{split}
\theta^{*}:=\underset{\theta\epsilon \Theta}{\mathrm{argmin}}\: \hat{J}(\theta).
\end{split}
\end{equation}
The objective function $\hat{J}(\theta)$ is available as expensive to evaluate the black box oracle on the real plant.

\vspace{-5 pt}
\section{Guided Bayesian Optimization}\label{sec:Guided_BO}
\begin{figure}[!t] 
\centering
\includegraphics[width=3.47in, trim=9cm 56cm 7.5cm 7cm, clip]{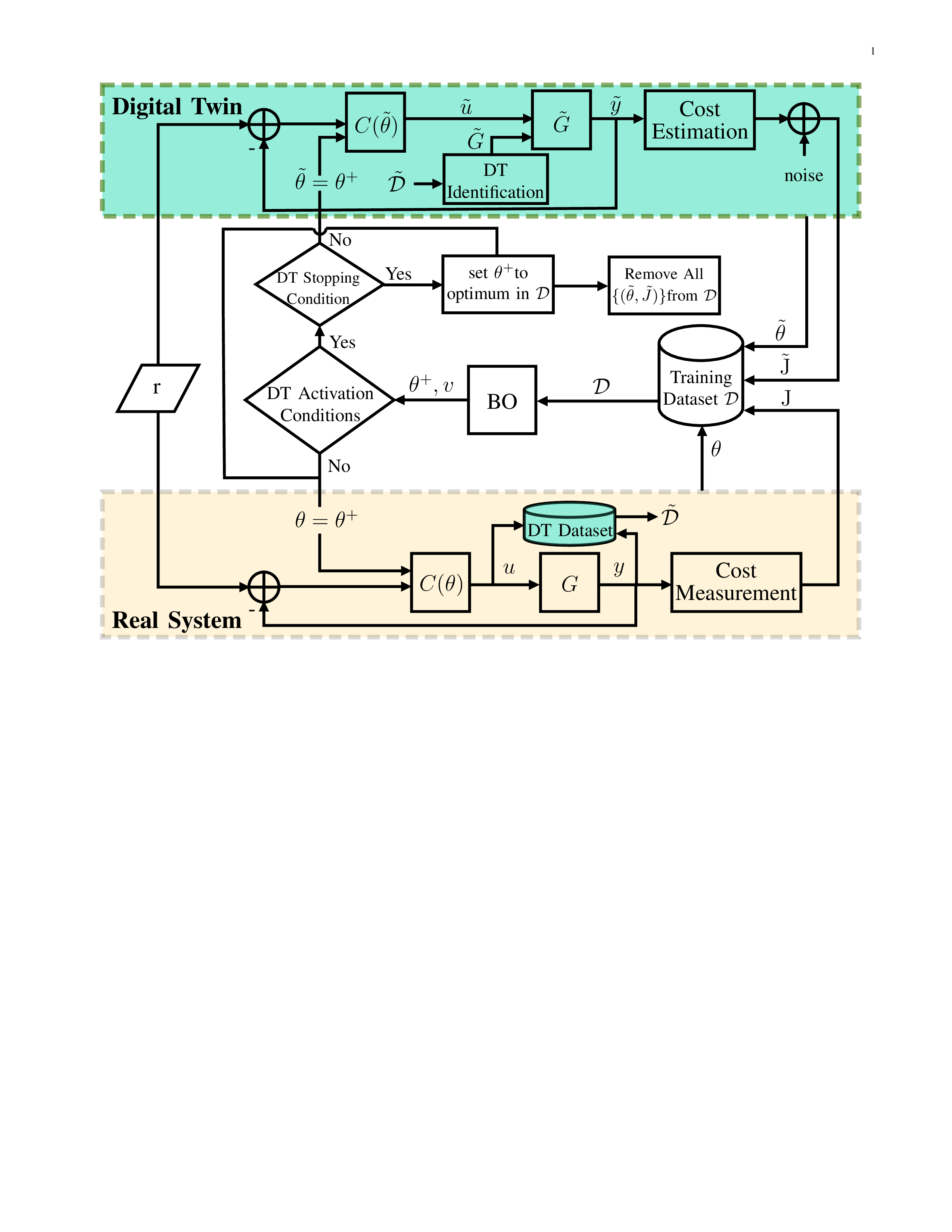}
\caption{Guided BO schematic representation. The digital twin of the closed-loop system is built using available data without additional operations on the real system.}
\label{fig:CLS_BO}
\end{figure}

The BO-based methods introduced in the literature inefficiently compress the data of each experiment to a single overall performance value.
For example, \cite{khosravi2020safety, VANNIEKERK2023103008, 9867298} maps only the output tracking error signal of the closed-loop system to a cost function, ignoring explicitly utilizing all the other internal signals.
However, intermediary data, such as control signals and output measurements, might be available and usually not explicitly exploited.
In this Section, we propose our novel \textit{guided} BO method shown in Fig. \ref{fig:CLS_BO}.
Our method utilizes the information flow in the closed-loop system to define a DT, improving the optimization and reducing operational costs. 
\vspace{-5 pt}
\subsection{Bayesian Optimization-based Controller Tuning}\label{subsec:BO}
Bayesian inference is a statistical inference method that uses Bayes' theorem to update the probability of a hypothesis as more data becomes available \cite{garnett_bayesoptbook_2023}.
BO can optimize complex ``black-box" objective functions using sparse observations \cite{wang2023recent, Maier2020}.
However, the computational disadvantage of the Bayesian inference approach is that it typically solves an intractable integration problem \cite{Rasmussen2004}.
Gaussian Process (GP) regression trains a stochastic model of data that provides predictive uncertainty and a tractable surrogate model for Bayesian inference. 




To build a prior distribution of the mean and covariance functions, BO requires an initial collection of control parameters $\Theta_{\text{init}}:= \left \{ \theta_{j} \in \Theta \: |\:  j=1,2,...,N_{0} \right \}$, where $N_{0} \geq 1$ is the number of parameters.
We use the Latin hypercube introduced by \cite{mckay2000comparison} as a space-filling sampling method to build $\Theta_{\text{init}}$.
So, each dimension of the feasible set $\Theta$ is divided into equally probable intervals where random control parameters are drawn.
This method selects the control parameters of the initial training data set independently and uniformly in each dimension of the feasible set. 

The BO algorithm maintains a pool of parameters and their corresponding performance values as a \textit{training data set} 
\begin{equation}\label{eq:D}
    \mathcal{D}:= \left \{ \bigl(\theta_{j}, J({\theta_{j}})\bigl)\:  |\:  \theta_{j} \in \Theta\ ;\ j=1,...,n ;\ n \geq N_{0} \right \},    
\end{equation}
in which we assume it initially contains $N_{0}$ data points.
Then, at every BO iteration, we estimate the cost function using Gaussian Process (GP) regression denoted by $\hat{J}$ based on $\mathcal{D}$.
More precisely, let $\mathcal{GP}(\mu,k)$ be the prior Gaussian process function with the prior mean and kernel functions  $\mu: \Theta\rightarrow \mathbb{R}$ and $k: \Theta \times \Theta \rightarrow \mathbb{R}_{\geq 0}$.
Consider parameter vectors $\mathrm{X}_{n}:=[\theta_{1},...,\theta_{n}]^\mathsf{T}$ and performance values $\mathrm{y}_{n}:=[J(\theta_{1}),...,J(\theta_{n})]^\mathsf{T}$, we write
\begin{equation}\label{eq:gpr}
\forall \theta \in \Theta: \hat{J}(\theta) \sim \ \mathcal{N}(\mu_{n}(\theta), v_{n}(\theta)),
\end{equation}
with mean and variance

\begin{align}\label{eq:gpr_mio1}
\mu_{n}(\theta)&:=\mu(\theta)+\mathrm{k}_{n}(\theta)^{\mathsf{T}}(\mathrm{K}_{n}+\sigma_{\epsilon}^{2}\mathbb{I})^{-1}(\mathrm{y}_{n}-\mu(\mathrm{X}_{n})),
\\\label{eq:gpr_v}
v_{n}(\theta)&:=k(\theta,\theta)-\mathrm{k}_{n}(\theta)^{\mathsf{T}}(\mathrm{K}_{n}+\sigma_{\epsilon}^{2}\mathbb{I})^{-1}\mathrm{k}_{n}(\theta),
\end{align}
where $\text{K}_{n}$ is defined as $[k(\theta_{\xi_{1}},\theta_{\xi_{2}})]\bigr\rvert_{\xi_{1},\xi_{2}=1}^{n}\in \mathbb{R}^{n \times n}$, $\sigma_{\epsilon}$ is the standard deviation of the additive noise to the predictive performance uncertainty, $\mathbb{I}$ is the identity matrix, vector $\mu(\mathrm{X}_{n})=[\mu(\theta_{1}),...,\mu(\theta_{n})]^\mathsf{T} \in \mathbb{R}^{n}$, and the vector $k_{n}(\theta) \in \mathbb{R}^{n}$ is equal to  $[k(\theta,\theta_{1}),...,k(\theta,\theta_{n})]^\mathsf{T}$.



The kernel function determines the relationship between neighboring points modeled by the GP.
Here, we utilize a Matérn kernel 5/2 that has demonstrated exemplary performance on noisy differentiable functions \cite{Rasmussen2004}.
We use a constant zero prior mean function without loss of generality.
We use available observations to optimize the prior mean and kernel by log marginal likelihood maximization.

To select query parameter vector $\theta_{n+1}$, we use an auxiliary \textit{acquisition function} to ensure a trade-off between exploring regions with high uncertainty and exploiting regions with a higher probability of finding the optimum gains.
We choose Expected Improvement (EI) \cite{gelbart2014bayesian} acquisition function with a closed-form formulation because it improves the optimization accuracy with a limited number of observations \cite{EI_Bull20112879, EI_Jones1998455, guidetti2021plasma}.
Let $a_{\text{EI},n}$ be the expected improvement acquisition function corresponding to the $n-$th iteration, defined as
\begin{equation}\label{eq:LCB}
a_{\text{EI},n}(\theta):=v_{n}(\theta)^{\frac{1}{2}}\bigl(z_{n}(\theta)\Phi(z_{n}(\theta))+\varphi(z_{n}(\theta))\bigl),
\end{equation}
with 
$z_{n}(\theta):=(\mu_{n}(\theta)-J_{n}^{+})v_{n}(\theta)^{-\frac{1}{2}}$ 
where $\mu_{n}(\theta)$ and $v_{n}(\theta)$ are the predictive mean and variance of the GP model, $J_{n}^{+}$ is the best-observed performance among the $n$ experiments in training data set $D$, $\Phi(\cdot)$ is standard normal cumulative distribution function, and $\varphi(\cdot)$ is the standard normal probability density function.

To optimize the acquisition function for the next parameter $\theta_{n+1}$, we exhaustively grid search with a fixed number of data within the feasible set $\Theta$:
\begin{equation}\label{eq:EI}
\theta_{n+1}:= \underset{\theta\epsilon \Theta}{\mathrm{argmax}}\: a_{\text{EI},n}(\theta).
\end{equation} 

Accordingly, we denote the next query point $\theta^{+}:=\theta_{n+1}$, measure the next performance value $J(\theta^{+})$, and update the \textit{BO training data set} as 
\begin{equation}\label{eq:D_training}
\mathcal{D}=
\mathcal{D}\cup \left \{ \bigl(\theta^{+},J(\theta^{+})\bigl) \right \}.
\end{equation} 
The computed sequence of $\theta_{1},\theta_{2},\theta_{3},...$ by this iterative optimization converges to the proper solution of \eqref{eq:J} \cite{frazier2018bayesian, gardner2014bayesian}.

The computational complexity of BO controller tuning at iteration $m$ is heavily influenced by inversion of the covariance matrix $\mathrm{K}_{n}$ in \eqref{eq:gpr_v} and is approximated by $O(m.n^{3})$ where $n$ is the size of training data set $\mathcal{D}$.
\vspace{-5 pt}



\subsection{Digital Twin}\label{subsec:DT}


In the BO-based controller tuning method, we already acquire informative data in each experiment so one could constantly learn a digital counterpart of the closed loop system.
Starting from an initial training data set with size $N_{0}$, we build the \textit{digital-twin data set} $\tilde{\mathcal{D}}$ at experiment $n$ as
\begin{equation}\label{eq:Ds}
\tilde{\mathcal{D}}:=\bigcup_{j=1}^{n} \bigcup_{i=1}^{\tilde{n}}\Big\{(u_{i,j},y_{i,j})\Big\},
\end{equation}
where  $\tilde{n}$ is the sampling length per each experiment, $u$ and $y$ are the real plant input and output, respectively.
For simplicity, we assume all the experiments have the same length $\tilde{n}$.


We estimate an approximate model of the real plant called \textit{DT plant} denoted by $\tilde{\mathrm{G}}$ based on $\tilde{\mathcal{D}}$.
Assume $\tilde{\mathrm{G}}$ is a linear time-invariant transfer function estimated by the refined instrumental variable method \cite{IV2015}.
This DT plant model is computationally cheap and does not require fast processing.
So our DT is the same closed-loop system but with a DT plant instead of the real plant.

The DT plant must have sufficient fidelity (see Section \ref{subsec:numerical_plant_fidelity}).
The fidelity of the DT plant model is estimated by Root Mean Square Error (RMSE)
\begin{equation}\label{eq:RMSE}
e_{\text{RMSE}}:=
\bigg[
\frac{1}{n \times \tilde{n}}\sum_{k=1}^{n}\sum_{j=1}^{\tilde{n}}{(y_{j}-\tilde{y}_{j})}^{2}_{k}
\bigg] 
^{\frac12},
\end{equation}
where $\tilde{y}$ is the output predicted by the DT plant.
\vspace{-5 pt}

\subsection{Guided Bayesian Optimization}\label{subsec:GBO_Algorithm}

Algorithm \ref{alg:guided} summarizes our guided BO method.
At each BO iteration using the physical plant $\mathrm{G}$ and given the query control parameters $\theta^{+}$, we collect the control signal $u$ and the plant output $y$ to build DT data set $\tilde{\mathcal{D}}$ and identify a DT plant model $\tilde{\mathrm{G}}$.
The DT plant is updated each time the performance of the closed-loop control system with the physical plant is measured.
Based on the GP model uncertainty, the guided BO algorithm gains more information from the updated DT.
More specifically, we activate the DT when two conditions 
\begin{subequations}\label{eq:DT_activation}
\begin{alignat}{2}
\label{eq:eta1}
\sqrt{v_{n}(\theta^{+})}& >&&\ \eta_{1}, \\ 
\label{eq:DT_rmse1}
e_{\text{RMSE}}& <&&\ \eta_{2}, 
\end{alignat}
\end{subequations}
are satisfied where $\eta_{1} \geq 0$ is the DT activation threshold, $v_{n}$ is the variance of the BO's posterior GP model calculated at the next query parameter vector $\theta^{+}$, $e_{\text{RMSE}}$ is RMSE defined in \eqref{eq:RMSE}, and $\eta_{2}$ is the RMSE threshold and is chosen based on the noise standard deviation as $\eta_2:=3\sigma_{\epsilon}$.
We estimate $\sigma_{\epsilon}^{2}$ by measuring the performance metrics in five experiments with identical controller gains in the feasible set.
When the DT is activated, we use the DT instead of the actual system to estimate the cost value denoted by $\tilde{\hat{J}}$ at $\tilde{\theta}:=\theta^{+}$.
We add a Gaussian noise with zero mean and $\sigma_{\epsilon}^{2}$ variance to the estimated performance to obtain $\tilde{J}:=\tilde{\hat{J}}+\epsilon$ where $\epsilon \sim \ \mathcal{N}(0, \sigma_{\epsilon}^{2})$.
This estimated performance $\tilde{J}$ is then added to the BO training data set 
in Line \ref{algLine:D_in_DT} of Algorithm \ref{alg:guided}.
In Section \ref{subsec:surrogate_frequency}, we numerically study the DT activation threshold $\eta_{1}$.

We continuously monitor the DT quality to maintain optimum controller parameters when the real plant changes. 
Each time we measure the performance on the real system $J$, we compare it with its estimated value by DT denoted by $\tilde{J}$.
Suppose that the system behavior has considerably changed.
In that case, we discard all previous data from the DT dataset.
More specifically, if the normalized difference between the measured and estimated costs is considerable, i.e., 
\begin{equation}\label{eq:reinit_DTdataset}
\frac{|J-\tilde{J}|}{J}>\tilde{\delta},
\end{equation}
where $\tilde{\delta}$ is DT re-initialization threshold, then we reinitialize DT dataset with the recent measurements as in Line \ref{algLine:reinit_D_tilde} of Algorithm \ref{algLine:reinit_D_tilde}.
If the system significantly changes, a new safe boundary for the optimization parameters must be determined, as demonstrated in \cite{SUKHIJA2023103922, ZAGOROWSKA202310107}, for example.

\begin{algorithm}[!t]
	\caption{Guided BO Controller Tuning Algorithm}\label{alg:guided}
	\begin{algorithmic}[1]
		\State \textbf{Input:} Set feasible set $\Theta$, initial control parameters $\Theta_{\text{init}}$,  weight vector $\mathrm{w}$, DT sampling length $\tilde{n}$, DT activation parameters $\eta_{1}$ and $\eta_{2}$, DT re-initialization parameter $\tilde{\delta}$, maximum experiments $n_{\text{max}}$, DT stopping threshold and consecutive iteration parameters $\eta_{3}$ and $\tilde{n}_{\text{EI}}$
        \State Initialize $\mathcal{D}$ and $\tilde{\mathcal{D}}$ using $\mathrm{G}$ to measure $u$, $y$, and $J(\theta), \forall \theta \in \Theta_{\text{init}}$
            \While{$n_{\text{max}}$ experiments on $\mathrm{G}$ is not reached}
            	\State Optimize $\mathcal{GP}(\mu,k)$ prior mean and kernel hyperparameters by minimizing negative log marginal likelihood
            	\State Estimate cost $\hat{J}$ as in \eqref{eq:gpr} 
            	\State Derive next query point $\theta^{+}$ solving \eqref{eq:EI}
                \If {DT activation criteria in \eqref{eq:DT_activation} is met}
                    \State Update DT plant model $\tilde{\mathrm{G}}$
                    \While{DT stopping criterion in \eqref{eq:DT_stop} is not met}
                        \State Use $\tilde{\mathrm{G}}$ to estimate cost $\tilde{\hat{J}}$ at $\tilde{\theta}=\theta^{+}$
                        \State Update BO training data: $\mathcal{D}\leftarrow \mathcal{D}\cup \left \{(\tilde{\theta},\tilde{J}) \right \}$\label{algLine:D_in_DT}
            			\State Optimize $\mathcal{GP}(\mu,k)$ prior hyperparameters
                        \State Estimate cost $\hat{J}$ as in \eqref{eq:gpr} 
                        \State Derive next query point $\theta^{+}$ solving \eqref{eq:EI}
                    \EndWhile
                    \State Set $\theta^{+}=\theta_{m} \in \mathcal{D} \text{ s.t. } \forall \theta \in \mathcal{D}: J(\theta_{m})<J(\theta) $ 
                    \label{alg:theta_plus}
                    \State Remove all DT data $(\tilde{\theta},\tilde{J})$ from $\mathcal{D}$ in \ref{eq:D} \label{algLine:remove_DT_data}
                \EndIf
            \State Use $\mathrm{G}$ to measure performance $J$ at $\theta=\theta^{+}$
            \State Update BO training data: $\mathcal{D}\leftarrow \mathcal{D}\cup \left \{(\theta,J) \right \}$
            \State Use $\tilde{\mathrm{G}}$ to estimate cost $\tilde{\hat{J}}$ at $\tilde{\theta}=\theta^{+}$
            \State Update DT data: $\tilde{\mathcal{D}}\leftarrow \tilde{\mathcal{D}}\cup \left \{ \bigcup_{i=1}^{\tilde{n}}(u_{i},y_{i}) \right \}$
            \If{DT re-initialization criterion in \eqref{eq:reinit_DTdataset} is met}
            \State Remove previous DT data: $\tilde{\mathcal{D}}\leftarrow  \left \{ \bigcup_{i=1}^{\tilde{n}}(u_{i},y_{i}) \right \}$\label{algLine:reinit_D_tilde}
            \EndIf
            \EndWhile
	\end{algorithmic} 
\end{algorithm}

An appropriate stopping criterion for the optimization process is necessary for the practical implementation of the algorithm.
We terminate the BO after a fixed number of iterations on the real system denoted by $n_{\max}$.
When using DT, we need a stopping condition that ensures prompt termination of the optimization. 
We use the stopping criterion that depends on the maximal expected improvement over previous iterations and the current expected improvement proposed in \cite{khosravi2020safety}.
Once we have activated the DT plant to predict the closed-loop response, we stop the BO iterations on DT  when the maximum number of iterations on the DT is reached or depending on the ratio between the last expected improvement and the maximal expected improvement realized so far.
This way, we avoid unnecessary iterations using the DT plant where we anticipate no further improvement.
Notably, we stop BO on DT when for $\tilde{n}_{{\text{EI}}}$ consecutive iterations on DT we meet one of the conditions 
\begin{subequations}\label{eq:DT_stop}
\begin{alignat}{2}
\label{eq:DT_stop_a}
a_{\text{EI},i}(\theta)\ &\leq&&\ \eta_{3}\: \underset{i-\tilde{n}_{EI}\leq j \leq i-1}{\mathrm{max}}\: a_{\text{{\text{EI}}},\text{j}}(\theta), \\ 
\label{eq:DT_stop_b}
i\ &>&&\ n_\text{max}, 
\end{alignat}
\end{subequations}
where $i$ is the number of BO iterations on DT \textit{after} last time DT is activated, $a_{\text{EI},i}(\theta)=a_{\text{EI}}(\theta^{+})$ and $\eta_{3}>0$ is a predefined threshold to stop BO iteration on DT as in \cite{Khosravi9324945}.
If the expected improvement does not significantly improve compared with the improvements in the last $\tilde{n}_{EI}$ iterations, then \eqref{eq:DT_stop_a} halts the optimization with DT.

As soon as we stop using DT, we update the query parameter vector $\theta^{+}$ with the optimum observed parameter vector in the BO training data set $\mathcal{D}$ as defined in line \ref{alg:theta_plus} of Algorithm \ref{alg:guided}.
Then, we remove all the DT data $(\tilde{\theta},\tilde{J})$ from $\mathcal{D}$.
We eventually proceed with the actual system to measure the overall performance at $\theta^{+}$.

The training data set size $n$ changes when we remove the DT data from $\mathcal{D}$ (see line \ref{algLine:remove_DT_data} of Algorithm \ref{alg:guided}).
Furthermore, at each iteration $m$ on the real plant (where $0 \leq m \leq n_\text{max}$), the DT can be activated and $i$ number of BO iterations can be performed on the DT where $i \leq n_\text{max}$.
So the computational complexity of the guided BO algorithm is bounded and dominated by a polynomial that depends on $i$, $m$, and $n$.

\vspace{-5 pt}
\section{System Modelling}\label{sec:System_Modelling}
Our guided BO is independent of the plant or parametric controller structure.
In this Section, we first specify the controller and the plant structure to numerically evaluate the guided BO algorithm in Section \ref{sec:Numerical_Analysis}.
Then, we explain our chosen performance metrics used to assess the system's overall performance in \eqref{eq:J}. 

We have a DC rotary motor with speed encoders equipped with a current converter and a cascaded speed controller.
As the two controllers operate on different time scales, we only consider the speed controller and aim to optimize its proportional and integral gains.
The DC motor with its encoder and the current converter is denoted by $\mathrm{G}$. 
\vspace{-10 pt}
\subsection{System Structure and Identification}
\label{subsec:System_Structure_Identification}
For the numerical studies in Section \ref{sec:Numerical_Analysis}, the real plant $\mathrm{G}$ is an identified model of the DC motor system around an operation point as a linear time-invariant (LTI) system.
We have a speed encoder with LabVIEW-based real-time interface \cite{KELLER2006177}.
The nominal encoder speed $\mathrm{v}$ is the number of pulses counted with sampling time $t=2$ ms.
We consider low and high reference encoder nominal speed inputs equal to $\mathrm{v}=1600$ and $\mathrm{v}=2000$, respectively.
Our encoder can count geometrically $N=2048$ pulses per revolution.
So the angular speed in radian per second can be calculated by $\omega=\frac{2\pi.\mathrm{v}}{N.t}$.

We measure the plant response to a given sinusoidal signal with various frequencies, a fixed amplitude, and a single DC value of $\bar{\Omega}=5522.40\ \text{rad/s}$.
Considering several excitation frequencies, the excitation signal has a fixed amplitude of $613.60\ \text{rad/s}$.
We measure the system output with a high sampling rate of $0.5$ ms and draw the bode diagram of the system response, which allows us to fit a second-order LTI transfer function visually.
We repeat each measurement two times to alleviate the measurement noise.
For A/D conversion, we sample 128 points evenly distributed in every period, introducing a constant phase shift in our bode diagram.
So, because we have $12$ different frequencies for the sinusoidal excitation signal, the sampling rate varies (see Table \ref{table:identification}).
Consequently, an input time delay of $2$ ms abstracts the time shift introduced by A/D conversion.
We address this delay by shifting the phase in the bode diagram to a fixed offset corresponding to the delay. 
Finally, $\mathrm{G}$ is written as
$\mathrm{G}(s):=\frac{L}{s^{2}+\tau_{1}s+\tau_{2}}\,\mathrm{e}^{-sl}$,
where Table \ref{table:identification} summarizes its identified parameters.
\vspace{-5 pt}
\setlength{\tabcolsep}{5.4pt}
\begin{table}[thb]
\centering
\setlength\extrarowheight{1.1pt}
\caption {\label{table:identification} Identification parameters of the linear time-invariant plant model around a specified operating point} 
\begin{tabular*}{0.45\textwidth}{@{\extracolsep{\fill}}*{4}{c}} \toprule
\multicolumn{2}{c}{Identified Plant Parameters} & \multicolumn{2}{c}{Identification Specifications}    \\  \cmidrule(lr{.5em}){1-2} \cmidrule(lr{.5em}){3-4}
 {Parameter} & {Value}   & {Parameter} & {Value}            \\    \midrule
 {$\tau_{1}$}     &  4.145         & {number of poles}         & 2         \\       
 {$\tau_{2}$}     &  4.199         & {number of frequencies}         & 12         \\       
 {$L$}      &  9.544         & {excitation amplitude [rad/s]}         & 613.60         \\       
 {$l$}      &   0.002        & {sampling rate [Hz]}         & 10-2000         \\     \bottomrule  
\end{tabular*}
\end{table}

\vspace{-25 pt}
\subsection{Controller Structure and Nominal Tuning} \label{subsec:nominal_tuning}
A PI speed controller\vspace{-5 pt}
\begin{equation}\label{eq:Cv}
\text{C}(s):=\mathrm{K}_{\text{p}}(1+\mathrm{K}_{\text{i}}\frac{1}{s}), 
\end{equation}
with $\mathrm{K}_{\text{p}}$ and $\mathrm{K}_{\text{i}}$ gains is defined.
For the controller parameter vector $\theta := [\mathrm{K}_{\text{p}},\mathrm{K}_{\text{i}}]$, we define a \textit{feasible set}\vspace{-5 pt}
\begin{equation}\label{eq:PI_gain_bounds}
\begin{split}
        \Theta := \Big\{[\mathrm{K}_{\text{p}},\mathrm{K}_{\text{i}}]  \:  \Big|  \: &\mathrm{K}_{\text{p}}\in [\mathrm{K}_{\text{p}_{\min}},\mathrm{K}_{\text{p}_{\max}}],\\
        &\mathrm{K}_{\text{i}}\in[\mathrm{K}_{\text{i}_{\min}},\mathrm{K}_{\text{i}_{\max}}] \Big\}.            
\end{split}
\end{equation}
We set the boundaries $\mathrm{K}_{\text{p}_{\min}},\mathrm{K}_{\text{p}_{\max}},\mathrm{K}_{\text{i}_{\min}},\mathrm{K}_{\text{i}_{\max}}$ equal to $0.11,1.10,0.87,2.08$, respectively, defining a rectangle around stable nominal controller parameters where the overshoot is bounded.

Classical tuning methods for the controller in \eqref{eq:Cv} involve numerical or graphical approaches using the Bode diagram \cite{ogata2010modern}.
We design a \textit{nominal controller} based on the real closed-loop system's phase margin and gain margin (PGM) \cite{FADALI200987}.
We analytically calculate the controller gains given PGM specifications with the simple approximations introduced by \cite{Ho1995} as follows\vspace{-5 pt}
\begin{align}
\label{eq:PGM:omega}
\omega_{p}& :=\frac{\text{A}_{\text{m}}\Phi_{\text{m}}+\frac{1}{2}{\pi}\text{A}_{\text{m}}(\text{A}_{\text{m}}-1)}{(\text{A}_{\text{m}}^2-1)l},
\\\label{eq:PGM:Kp}
\mathrm{K}_{\text{p}}& :=\frac{\omega_{p}\tau}{\text{A}_{\text{m}}L},
\\\label{eq:PGM:Kv}
\mathrm{K}_{\text{i}}& :=2\omega_{p}-\frac{4\omega_{p}^2l}{\pi}+\frac{1}{\tau},
\end{align}
where $\omega_{p}$ is the approximated phase crossover frequency.
We choose two value pairs for phase and gain margins to estimate nominal controllers that assure the under-damped behavior of the closed loop system.
Table \ref{table:nominal_ctl_specs} lists the resulting nominal controller gains and their PGM.

\begin{table}[!h]
\centering
\setlength\extrarowheight{1.1pt}
\caption {\label{table:nominal_ctl_specs} Specifications of Nominal Controllers} 
\begin{tabular*}{0.4\textwidth}{@{\extracolsep{\fill}}*{6}{c}} \toprule
{Controller name} & {$\Phi_{\text{m}}$} & {$\text{A}_{\text{m}}$} & {$\mathrm{K}_{\text{p}}$} & {$\mathrm{K}_{\text{i}}$} \\ \midrule
    {First Nominal PGM}     &  60{°} & 46 {dB} &  0.85 & 1.07     \\
    {Second Nominal PGM}     &  75{°} & 46 {dB} &  0.86 & 0.89 \\  \bottomrule

\end{tabular*}
\end{table}
\vspace{-25 pt}
\subsection{Performance Metrics}\label{subsec:Performance_metrics}
Following \cite{Khosravi9324945}, we define four performance indicators namely percentage overshoot ($\zeta$), settling time ($T_{\text{s}}$), rise time ($T_{\text{r}}$) and the integral of time-weighted absolute error ($e_{\text{ITAE}}$).
\subsubsection{Percentage overshoot}
If $y(t)$ is the step response of a continuous-time system, the percentage overshoot at each transition from low initial state $r_{1}$ to the high final reference input $r_{2}$ is defined such that \vspace{-5 pt}
\begin{equation}\label{eq:zeta}
\zeta:=100 \times \text{max}(0,\frac{M_{\text{p}}}{\Delta r}),
\end{equation}
where $y_{\text{max}}$ is the maximum response in the transition period, $\Delta r:=r_{2}-r_{1}$ is the step height, $M_{\text{p}}:=y_{\text{max}}-y(T)$ is the peak overshoot, and $T$ is the final time.
Percentage overshoot presents the stability degree and damping in the system \cite{samur2012performance}.

\subsubsection{Settling time} This is the time that the system response takes to converge and stay within an error band, i.e., $2\%$ of the step height, from the final value \cite{Tay1998}.
Settling time indicates a closed-loop system's convergence time to reach its steady state.
The settling time $T_{\text{s}}$ is defined as
\begin{equation}\label{eq:Ts}
\forall t\geq T_{\text{s}}: \left | y(t)-r_{2} \right |\leq \frac{2}{100}\times{\Delta r}.
\end{equation}
\vspace{-5 pt}
\subsubsection{Rise time} 
\begin{figure}[!h] 
\centering
\includegraphics[width=3.45in, trim=4.15cm 0.6cm 4cm 0.75cm, clip]{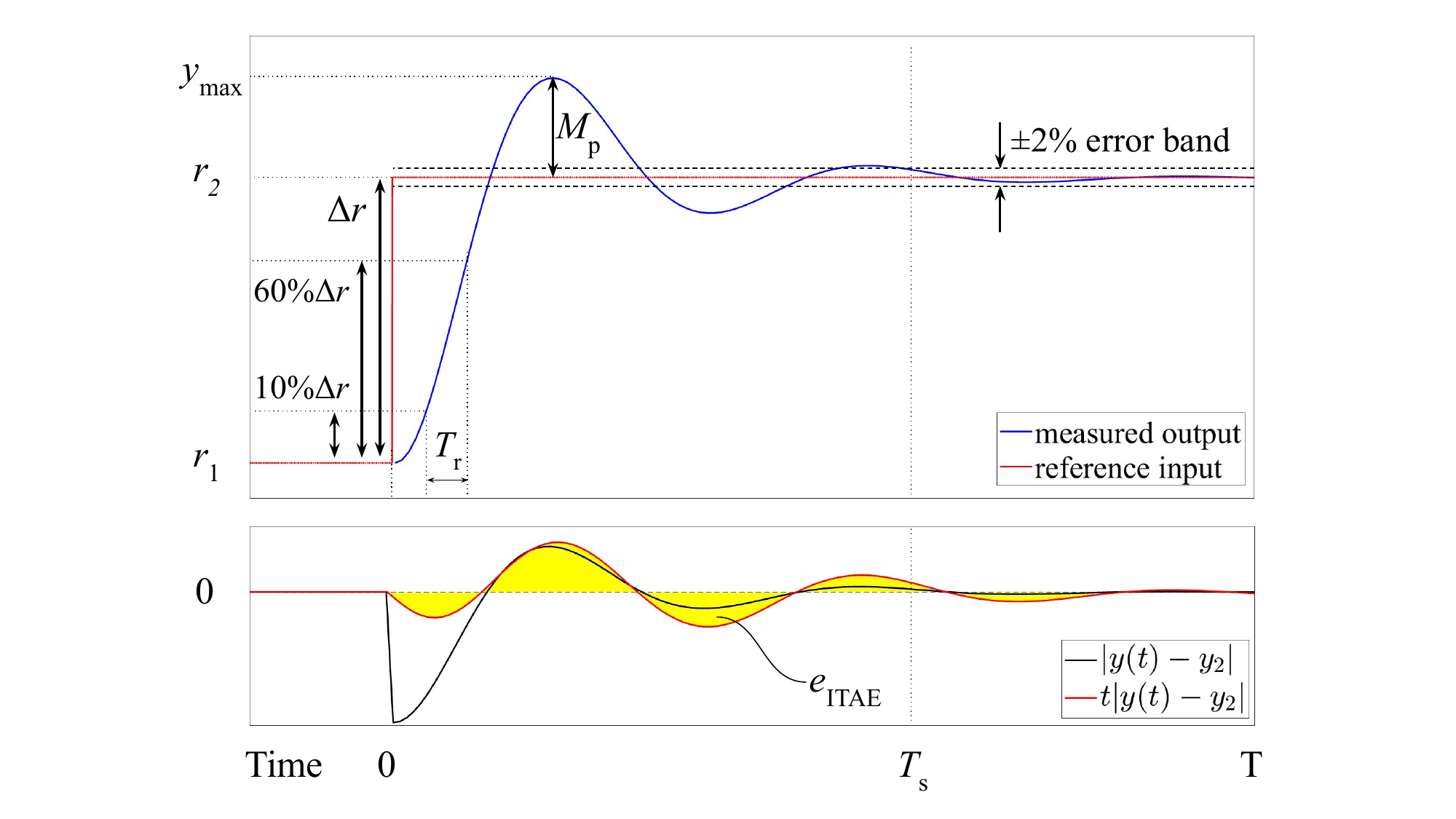}
\caption{Details of the performance metrics used to define the overall cost function}
\label{fig:perf_metrics}
\end{figure}

$T_{\text{r}}$ indicates the time for the step response to rise from $10\%$ to $60\%$ of the way from $r_{1}$ to $r_{2}$ (see Fig. \ref{fig:perf_metrics}).
So, a shorter rise time indicates a rapid response to the changes in the input signal. 
\vspace{-5 pt}

\vspace{-8 pt}

\subsubsection{Integral of time-weighted absolute error} Integral of time-weighted absolute error (ITAE) is the absolute time-weighted area of the step response error from the reference signal.
This performance index reduces system overshoot and oscillations \cite{4783932}, allowing a smoother response.
We write ITAE as
\begin{equation}\label{eq:ITAE}
e_{\text{ITAE}}:= \int_{0}^{T}t\left | y(t)-r_{2}) \right |dt ,
\end{equation}
where the final time $T \geq T_{\text{s}}$ is a fixed period.
Fig. \ref{fig:perf_metrics} visualizes details of our selected performance metrics.

According to the overall cost function in \eqref{eq:J}, a weighted sum of four performance metrics is defined.
The system performs well if its step response rises fast enough with minimum overshoot and fluctuations to a stable bandwidth.
We can write  
\begin{equation}\label{eq:J_used}
J(\theta)=w_{1}\zeta + w_{2}T_{\text{s}}+w_{3}T_{\text{r}}+w_{4}e_{\text{ITAE}},
\end{equation} such that $\sum_{i=1}^{i=4}w_{i}=1$ and $w_{i}>0$ where $w_{i}$ are the normalized weights of our selected individual performance metrics.
Different weights can be used to attain different trade-offs between the performance metrics; in the experiments below, we measure each performance metric using $5$ controller gains evenly spanning the entire feasible set $\Theta$, then take the average of each metric as a normalization weight.

We adjust the weights based on the relative significance of each metric.
For instance, we put more relevant gravity on minimizing overshoot related to performance and stability.
The normalized weights $w_{1},w_{2},w_{3},w_{4}$ are respectively selected as $0.44, 0.22, 0.22, 0.11$, summing up to unit value.

The following Section will analyze our guided BO properties on the identified plant $\mathrm{G}$.

\vspace{-5 pt}
\section{Numerical Analysis}\label{sec:Numerical_Analysis}
We numerically study the guided BO properties using the controller \eqref{eq:Cv} and plant $\mathrm{G}$ comparing with the nominal controller tuning.
Computations are performed by the Euler high-performance cloud-based cluster of ETH Zurich \cite{euler_eth}.
We use $8$ AMD EPYC 7742 CPUS cores with a clock speed of $2.25$ GHz nominal and $3.4$ GHz peak and $3072$ MB of memory. 

The reference input $r$ of the closed-loop system equals the angular encoder speed $\omega$ in radian per second. 
We consider low and high reference encoder speed inputs equal to $r_{1}=245$ rad/s and $r_{2}=306$ rad/s corresponding to the nominal encoder speeds of $\mathrm{v}=1600$ and $\mathrm{v}=2000$, respectively.
We model the output measurement noise by an additive Gaussian noise with zero mean and standard deviation $\sigma_{\epsilon}=0.03$.

We calculate the ground truth speed controller parameters $\theta^{*}$ with an exhaustive grid search in the feasible set $\Theta$ such that the cost in \eqref{eq:J_used} is minimized.
The \textit{optimality ratio} $\phi$ is the ratio between the cost and the ground truth optimum cost value $J(\theta^{*})$.
We calculate once the $\theta^{*} \in \Theta$ with a dense grid search where the minimum cost on the numerical system is retrieved (see Table \ref{table:numerical_gains}).
The optimality ratios of our nominal controllers in Table \ref{table:nominal_ctl_specs} are $\phi=1.71$ and $\phi=1.49$, respectively. 


We will analyze the DT plant fidelity, the proportion of BO iterations on DT, and the number of data in the initial training data set $\mathcal{D}$ on the identified plant.
To build the DT data set in \eqref{eq:Ds}, for simplicity, we fix the DT re-initialization threshold $\tilde{\delta}=2$ and the sampling length $\tilde{n}=100$ in a fixed period $T$ for all experiments.
We summarize the parameter values used in our guided BO algorithm in Table \ref{table:parameter_values}.

\setlength{\tabcolsep}{4.4pt}
\begin{table}[!h]
\centering
\caption {\label{table:parameter_values} Parameter values of our guided BO used in corresponding sections, including the range for varied parameters} 
\begin{tabular*}{0.47\textwidth}{@{\extracolsep{\fill}}*{8}{c}} \toprule
{Section} & {System} & {$\eta_{1}$} & {$\eta_{2}$} & {$\eta_{3}$} & {$N_{0}$} & {$\tilde{n}_{\text{EI}}$}& {$n_{\text{max}}$} \\ \midrule
    {\ref{subsec:numerical_plant_fidelity}}    & numerical  & 0-$\infty$ & 0.09-2.5 & $\infty$ & 1  & 5-40 & 50\\
    {\ref{subsec:initial_training_set_size}}   & numerical  &  0-$\infty$ & 0.09 & $\infty$ & 1-10  &  5 & 50\\ 
    {\ref{subsec:surrogate_frequency}}         & numerical  &  $10^{-6}$-20 & 0.09 & 0.2 & 1  & 3 & 50\\
    {\ref{subsec:performance_evaluation}}      & numerical  &  3 & 0.09 & 0.2 & 1  & 3 & 35\\
    {\ref{subsec:rotary_motor_system}}         & rotary motor & 3 & 0.09 & 0.2 & 1  & 3 & 40\\
    {\ref{subsec:linear_servo_motor_system}}   & linear motor & 3 & 0.003 & 0.2 & 1  & 3 & 25\\ \bottomrule
\end{tabular*}
\end{table}

\vspace{-15 pt}
\subsection{Digital Twin Fidelity}\label{subsec:numerical_plant_fidelity}
We identify two DT plants using second-order and fifth-order LTI models to assess the DT fidelity on the guided BO performance, as described in Section \ref{subsec:DT}.
We sample uniform data from the actual system's step response over a fixed period $T=5$ s, which is longer than the nominal closed-loop settling time.
We compare the fidelity of these two DT plant models with RMSE in \eqref{eq:RMSE} at the beginning of optimization using the DT data set $\tilde{D}$.
Second order DT plant is a \textit{high-fidelity} model with RMSE of $e_{\text{RMSE}}=0.01$; whereas the fifth-order model called \textit{low-fidelity} DT plant has a higher RMSE equal to $e_{\text{RMSE}}=2.41$.

We run a Monte Carlo analysis of the guided BO algorithm using both DTs with $100$ batches of experiments.
Each batch includes $n_{\text{max}}=50$ experiments on the real plant starting from an initial data set with a single data $N_{0}=1$.
The controller gains of the initial data set are randomly selected inside the feasible set.
So, the initial data sets of the batches are different from each other.
Starting from the initial training data set, we set $\eta_{1}=0$ in \eqref{eq:eta1} and perform BO on the DT plant $\tilde{n}_{\text{EI}}$ times.
Each time we activate DT, we continue BO iterations on DT up to a fixed number of iterations $\tilde{n}_{\text{EI}}$.
Next, we deactivate the DT by setting $\eta_{1}=\inf$, and continue $N_{\mathrm{G}}$ iterations on the real plant until we activate the DT plant by setting $\eta_{1}=0$.
This way, we activate DT periodically after every $N_{\mathrm{G}}$ iteration on the real plant $\mathrm{G}$.
Here we set $\eta_{2}=0.09$ and $\eta_{2}=2.5$ when using high- and low-fidelity DT plants, respectively.

\begin{table}[h!]
\centering
\setlength\extrarowheight{1.1pt}
\caption {Monte Carlo results with $100$ batches of $50$ experiments on the real plant, each starting from a different initial data set $\mathcal{D}$. Columns $3-5$ report the rounded average number of experiments over the real plant required to outperform the nominal optimality ratio $\phi=1.49$.}
\label{table:G2_sensitivity}
\begin{tabular*}{0.48\textwidth}{@{\extracolsep{\fill}}*{5}{c}} \toprule
\multirow{2}{*}{$\tilde{n}_{\text{EI}}$} & \multirow{2}{*}{$N_{\mathrm{G}}$} & BO &  \multicolumn{2}{c}{Guided BO}   \\ \cmidrule{3-3} \cmidrule{4-5}
    &   &   no DT   &   high-fidelity DT   & low-fidelity DT                                  
\\    \midrule
 5      &  1    & 11     & 7     & $>$50     \\       
 5      &  3    & 11     & 9     & 16     \\       
 5      &  5    & 11     & 7     & 16     \\       
 5      &  7    & 11     & 8     & 15     \\       
 10      &  1    & 11     & 6     & 16     \\       
 10      &  3    & 11     & 8     & 16     \\       
 10      &  5    & 11     & 7     & 16     \\       
 20      &  1    & 11     & 7     & 15     \\       
 40      &  1    & 11     & 7     & 16     \\     \bottomrule  

\end{tabular*}
\end{table}


Results on Table \ref{table:G2_sensitivity} show that with the same number and frequency of DT integration, the guided BO with high-fidelity DT converges to a desired optimality ratio sooner than BO.
Thus, the DT directly affects the convergence of the guided BO method. 

\vspace{-5 pt}
\subsection{Initial Training Set Size}\label{subsec:initial_training_set_size}
We study how the number of initial data $N_{0}$ measured on the real plant contributes to the performance.
We perform Monte Carlo analysis with $100$ batches of experiments.
For each batch, we build the initial training data set $\mathcal{D}$ by selecting $N_{0}$ number of controller vectors $\theta$ using a Latin hyper-cube sampling method inside the feasible set $\Theta$.
According to Table \ref{table:N0_sensitivity}, for a small $N_{0}$, the guided BO converges on average $3$ experiments faster than BO.
In this case, DT compensates for insufficient initial information.
The DT benefit for the guided BO diminishes when the initial training set contains more data. 

\begin{table}[h!]
\centering
\setlength\extrarowheight{1.1pt}
\caption {Monte Carlo results on $100$ batches of experiments with different initial data set size $N_{0}$. Each row is the rounded average number of iterations on the real plant required to outperform the nominal controller.}\label{table:N0_sensitivity}
\begin{tabular*}{0.4\textwidth}{@{\extracolsep{\fill}}*{5}{c}} \toprule
\multirow{2}{*}{$N_0$} & \multicolumn{2}{c}{$\phi=1.71$} & \multicolumn{2}{c}{$\phi=1.49$}    \\ \cmidrule{2-3} \cmidrule{4-5}
        &  Guided BO   & BO                                                            
	    &  Guided BO   & BO                                            \\    \midrule
 1      &  5            & 9         & 13         & 17         \\       
 5      &  4            & 5         & 14         & 14         \\       
 10     &  5            & 6         & 17         & 19         \\     \bottomrule  

\end{tabular*}
\end{table}

\vspace{-14pt}
\subsection{Digital Twin Activation Threshold}\label{subsec:surrogate_frequency}

\begin{figure}[!h]
\centering
\subfloat[]
{	\label{subfig:sensivity_eta1_convergence}
	\includegraphics[width=0.38\textwidth,trim=0cm  0cm 0cm  0cm, clip]{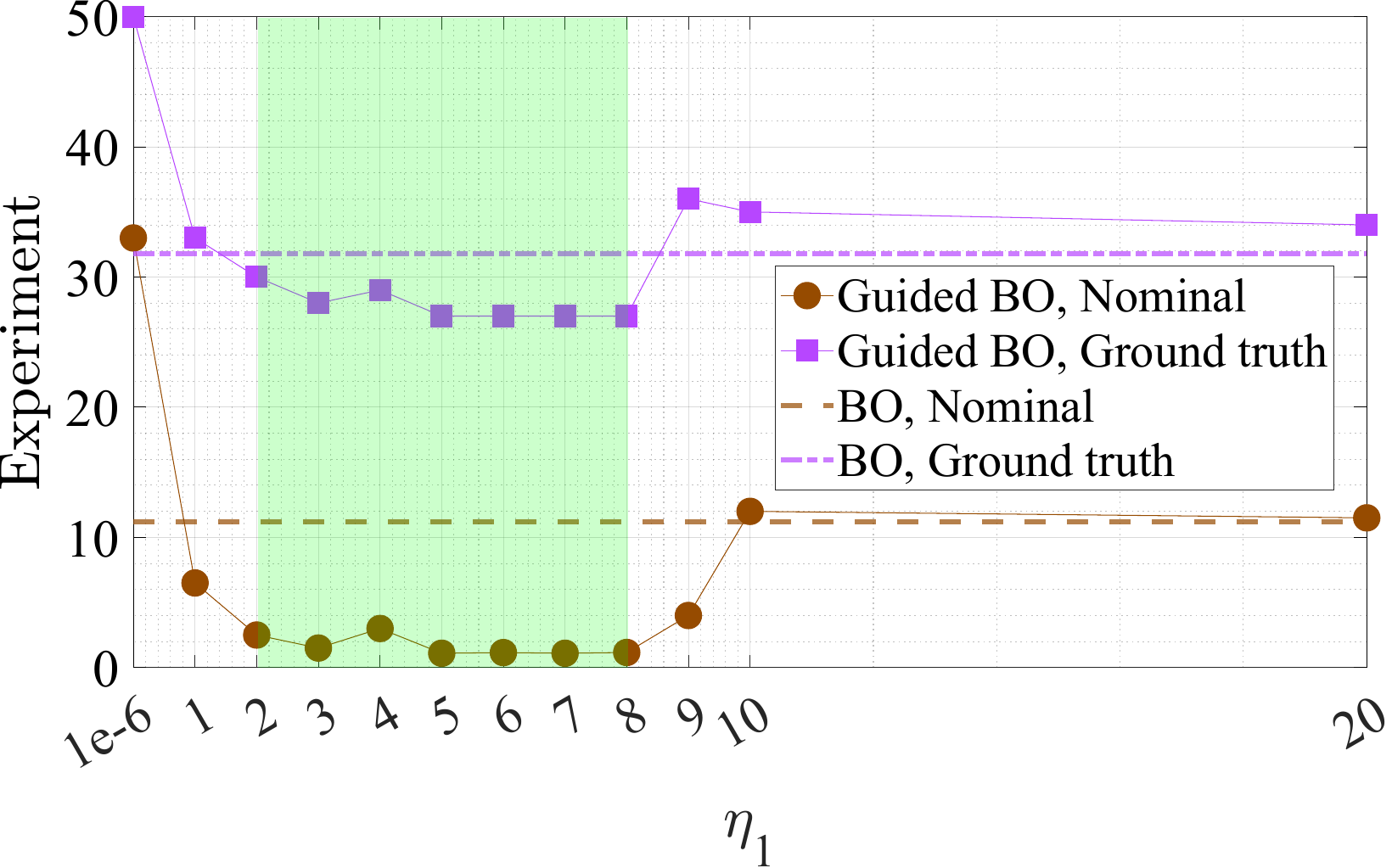}}

\subfloat[]{
	\label{subfig:sensivity_eta1_e_n}
	\includegraphics[width=0.43\textwidth,trim=0cm  0cm 0cm  0cm, clip]{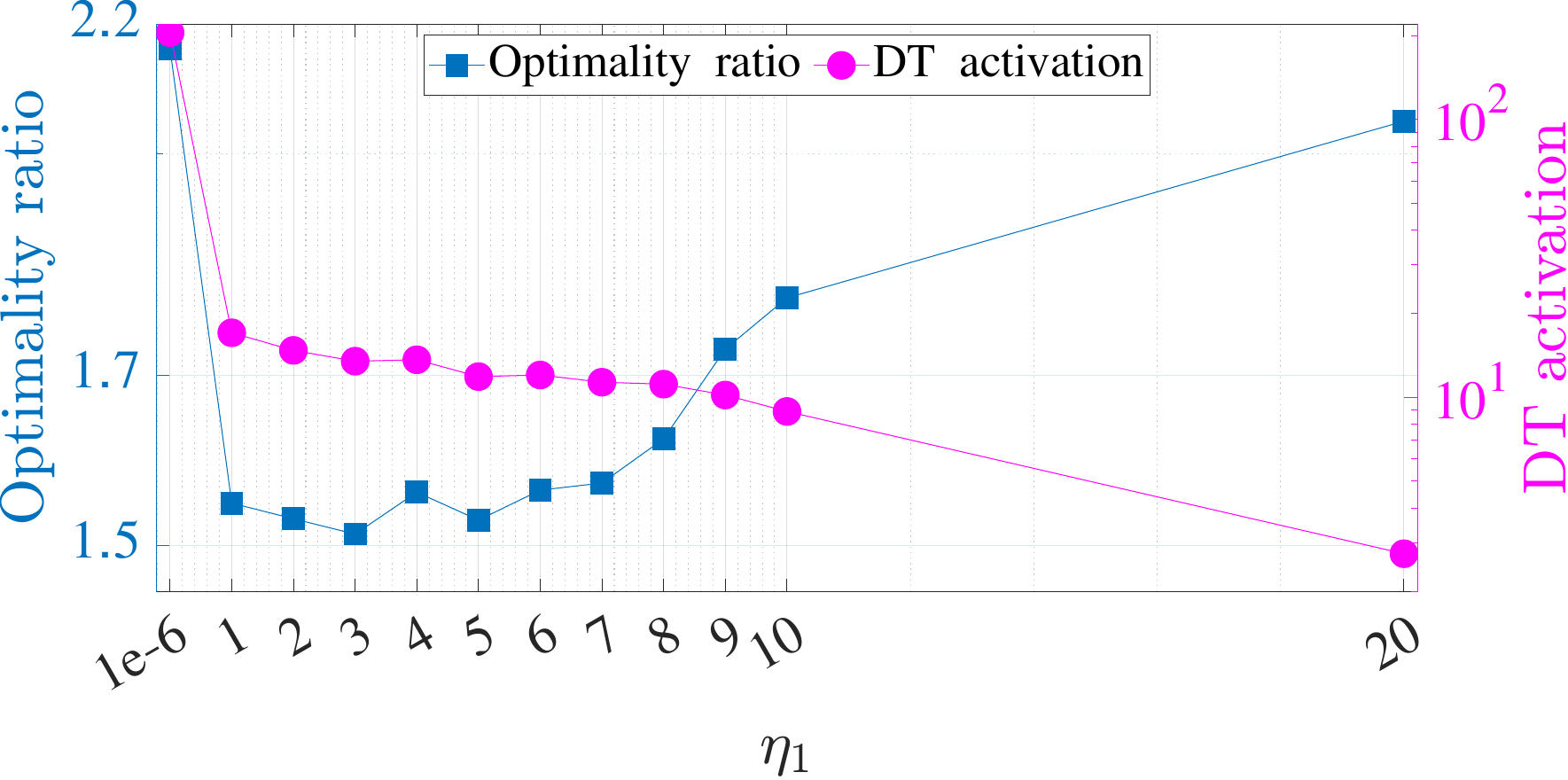}} 
\caption{\label{fig:eta1-sensivity} Results for different activation threshold $\eta_{1}$ of DT in guided BO method. Each data point averages over $100$ batches, each including $50$ experiments on the real plant.
\textbf{(top)} The required number of experiments on the real system to outperform the nominal controller performance or to converge to the ground truth performance $J(\theta^{*})$. The shaded area indicates $\eta_{1}$ range where guided BO converges faster to the ground truth performance than BO.
\textbf{(bottom)} The \textit{left axis} is the average optimality ratio. The \textit{right axis} shows the average number of DT activations in each batch.}
\end{figure}

We determine the DT activation threshold $\eta_{1}$ in \eqref{eq:eta1}.
At each threshold $\eta_{1}$, we repeat the guided BO algorithm $100$ times using different initial data sets.
Fig. \ref{fig:eta1-sensivity} shows that guided BO tunes the controller gains faster than BO when the DT is sufficiently activated, allowing the average optimality ratio of the measured performance to be low enough. 
More precisely, we observe in Fig. \ref{subfig:sensivity_eta1_convergence} that for guided BO, choosing a DT activation threshold in the range $\eta_{1} \in [2,8]$ (corresponding to DT activation around $12$ times on average in each batch) results in improved convergence of approximately $9$ experiments to the nominal performance compared to the BO method.
We also observe that with large $\eta_{1}=20$, the guided BO approaches the BO's performance indicated by dashed lines.
\vspace{-2 mm}

\vspace{-5 mm}
Fig. \ref{fig:eta1-sensivity} indicates the correlation between the total DT activations and the average optimality ratio measured on the real plant using various $\eta_{1}$ thresholds.
For the DT activation threshold $\eta_{1} \geq 5$ in \ref{eq:eta1}, increasing $\eta_1$ decreases the number of times DT guides the optimizer and increases the average optimality ratio measured on the real plant.
However, the cost estimated by DT is inaccurate, and over-activating the DT results in lower performance.
That is why using small $\eta_{1}=1e-6$ results in the excessive activation of the DT, in which the average optimality ratio also increases to $2.2$.
Therefore, choosing the DT activation threshold in the range $\eta_{1} \in [2,8]$ provides a better trade-off between the DT activation and the guided BO performance.
\vspace{-8 pt}
\subsection{Performance Evaluation}\label{subsec:performance_evaluation}

\begin{figure}[!h] 
\centering
\includegraphics[width=0.4\textwidth]{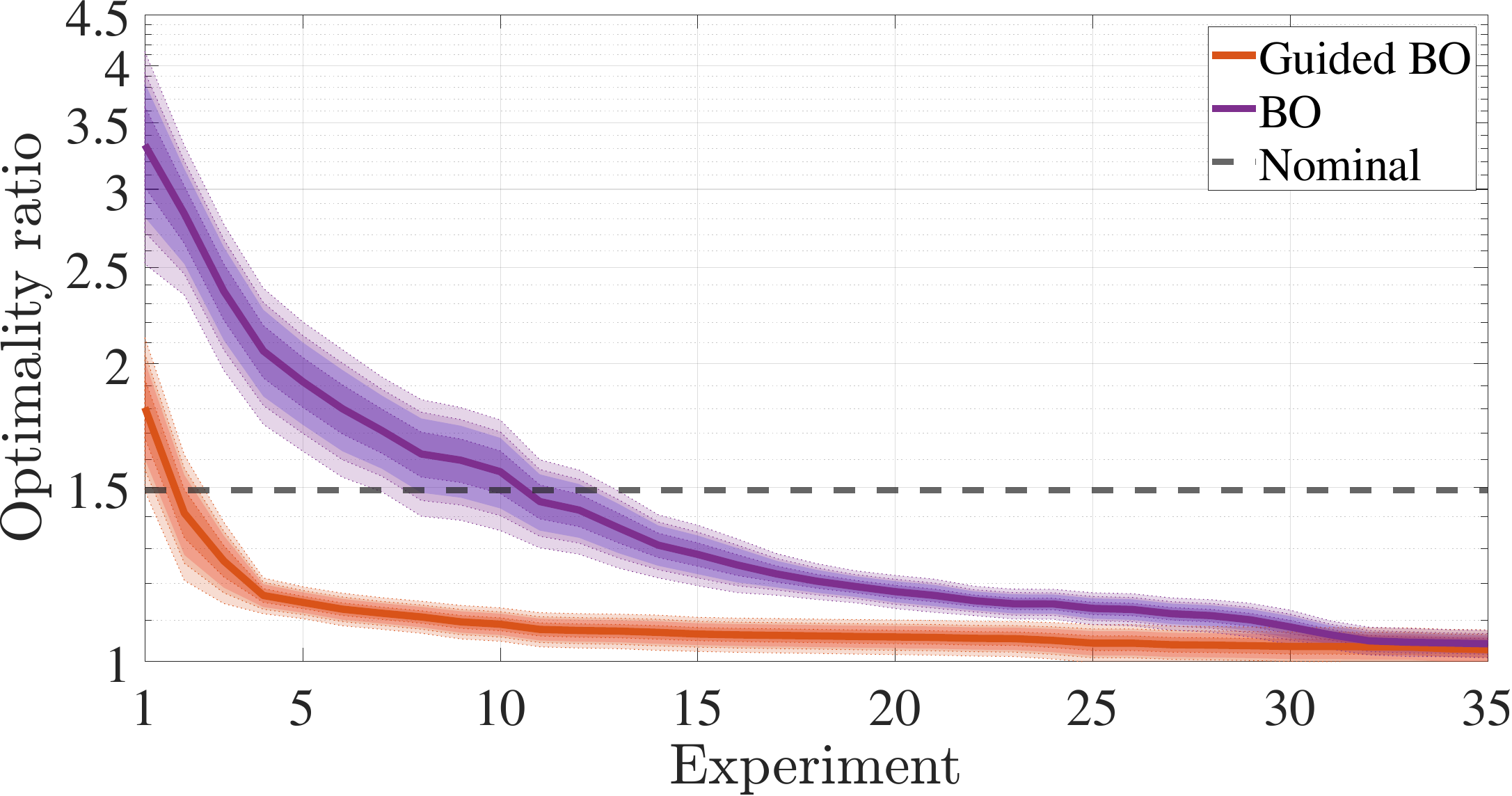}
\caption{Minimum observed optimality ratio up to number of BO experiments on the real plant. The thick line is the average over $100$ batches, and the shaded area shows the $99\%$, $95\%$, $90\%$, and $68\%$ confidence intervals.}
\label{fig:numerical_min}
\end{figure}

A Monte Carlo study is performed with $100$ batches of experiments.
Table \ref{table:parameter_values} specifies the guided BO algorithm parameter values.
According to \ref{subsec:surrogate_frequency}, the parameter $\eta_1=3$ is chosen to provide us with an optimum trade-off between DT activation and the convergence speed of the algorithm.
$\eta_{2}=0.09$ is chosen concerning three times the system noise boundary compared to the expected fidelity of the DT through RMSE error.
We set the stopping threshold $\eta_{3}=0.2$ and $\tilde{n}_{\text{EI}}=3$ to avoid unnecessary optimization on the DT.
The minimum number of initial data $N_{0}=1$ is chosen to provide the GP model with the least prior information when the guided BO is shown to improve the optimization further compared to BO as revealed in section \ref{subsec:initial_training_set_size}.
The results are shown in Fig. \ref{fig:numerical_min}.
Guided BO with high-fidelity DT outperforms the nominal controller on average after approximately $2$ experiments, whereas BO requires approximately $12$ experiments. 

\begin{figure}[!h]
\centering
{	\label{subfig:KpKiCost_numerical}
	\includegraphics[width=0.42\textwidth,trim=2.1cm  0cm 0cm  1cm, clip]{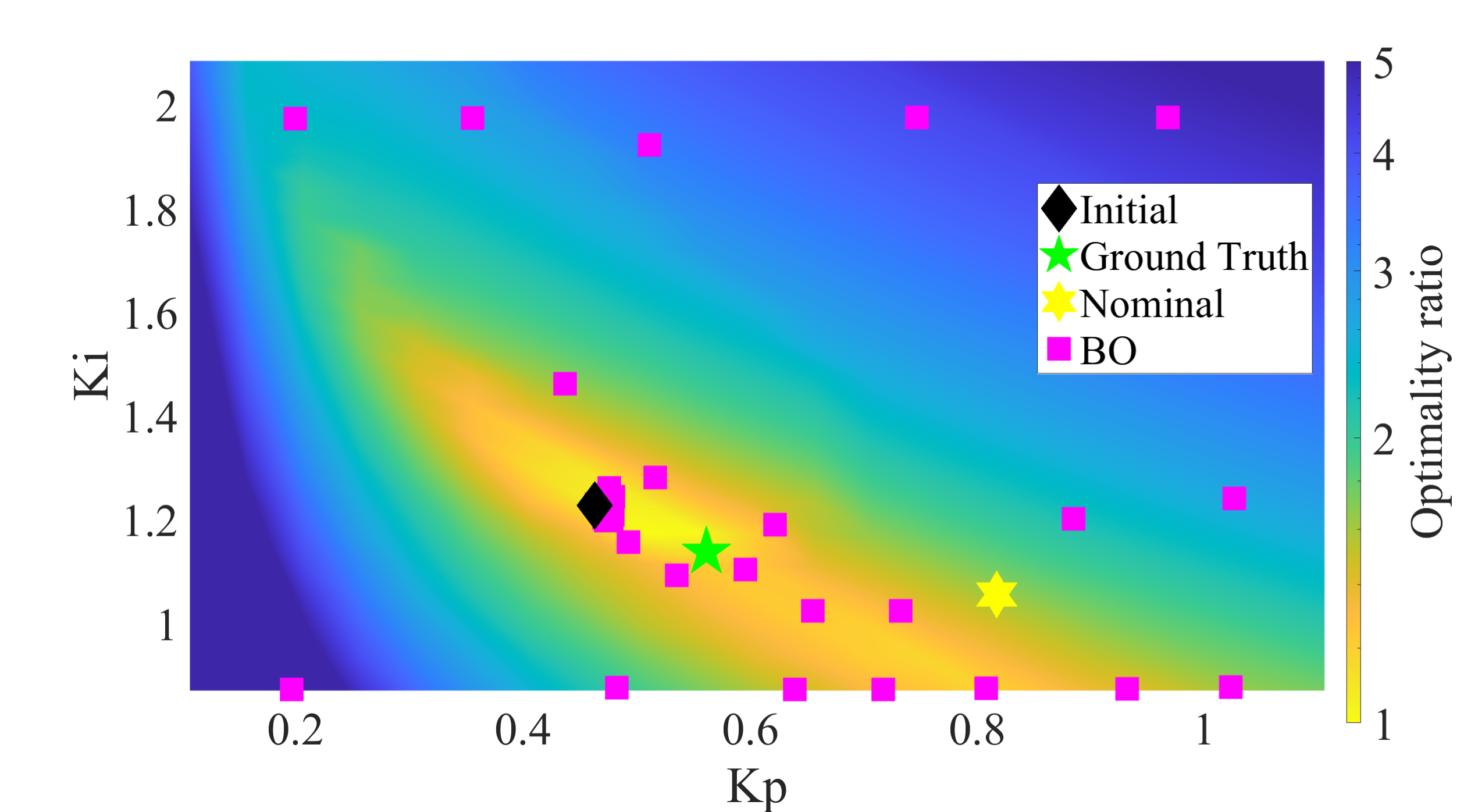}}
{
	\label{subfig:KpKiCost_DTs}
	\includegraphics[width=0.42\textwidth,trim=2.1cm  0cm 0cm  1cm, clip]{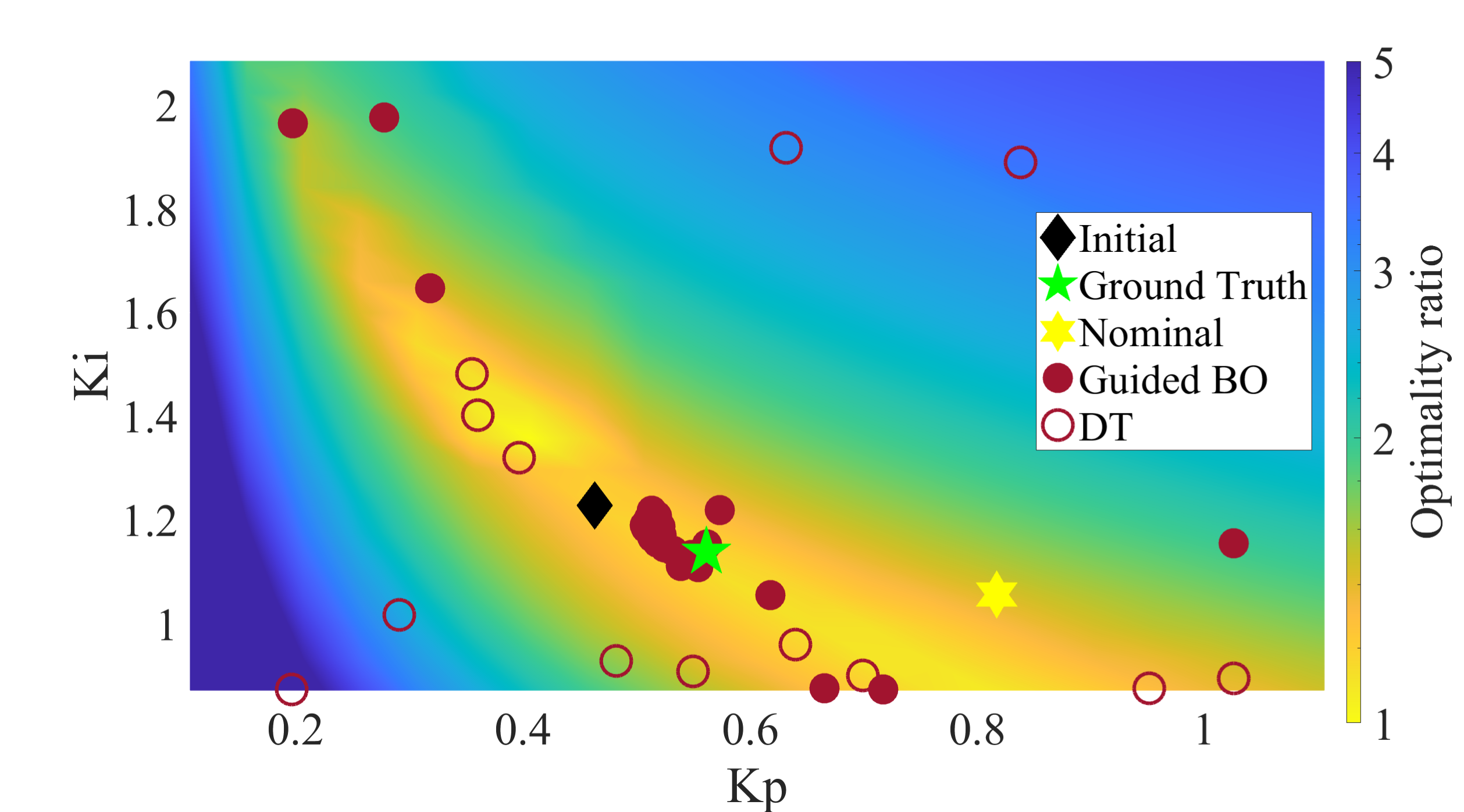}} 
\caption{\label{fig:KpKiCost}Control parameters proposed by the acquisition function to measure performance in one batch of $35$ iterations in guided BO and BO starting from the same initial data set.
\textbf{(top)} Colormap shows \textit{true cost} surface using the real plant.
\textbf{(bottom)} Colormap shows the \textit{estimated cost} by DT plant after $34$ iterations on the numerical system. Hollow circles are the gains where DT estimates the performance.}
\end{figure}

\vspace{-5 pt}
We compare the evaluated controller parameters of guided BO and BO in a batch of experiments starting from the same initial controller parameters $\Theta_{\text{init}}$.
In Fig. \ref{fig:KpKiCost}, we visualize one batch of our Monte Carlo analysis consisting of $35$ control parameters proposed by each optimizer to measure the performance with the real plant.
The true cost surface is presented with a colormap in Fig. \ref{fig:KpKiCost} which is a non-convex function of the controller parameters.
This figure shows that the BO (magenta squares) requires exploring the feasible set.
Fig. \ref{fig:KpKiCost} (bottom panel) depicts the cost estimated by DT after $34$ iterations on the real plant where the filled circles represent the measured parameters by guided BO on the real plant, and the hollow circles show the DT evaluations.
We see that DT retrieves the true cost.
Thus, the guided BO exploits the optimum region, avoiding the regions with higher costs.

\begin{figure}[!h] 
\centering
\includegraphics[width=0.4\textwidth]{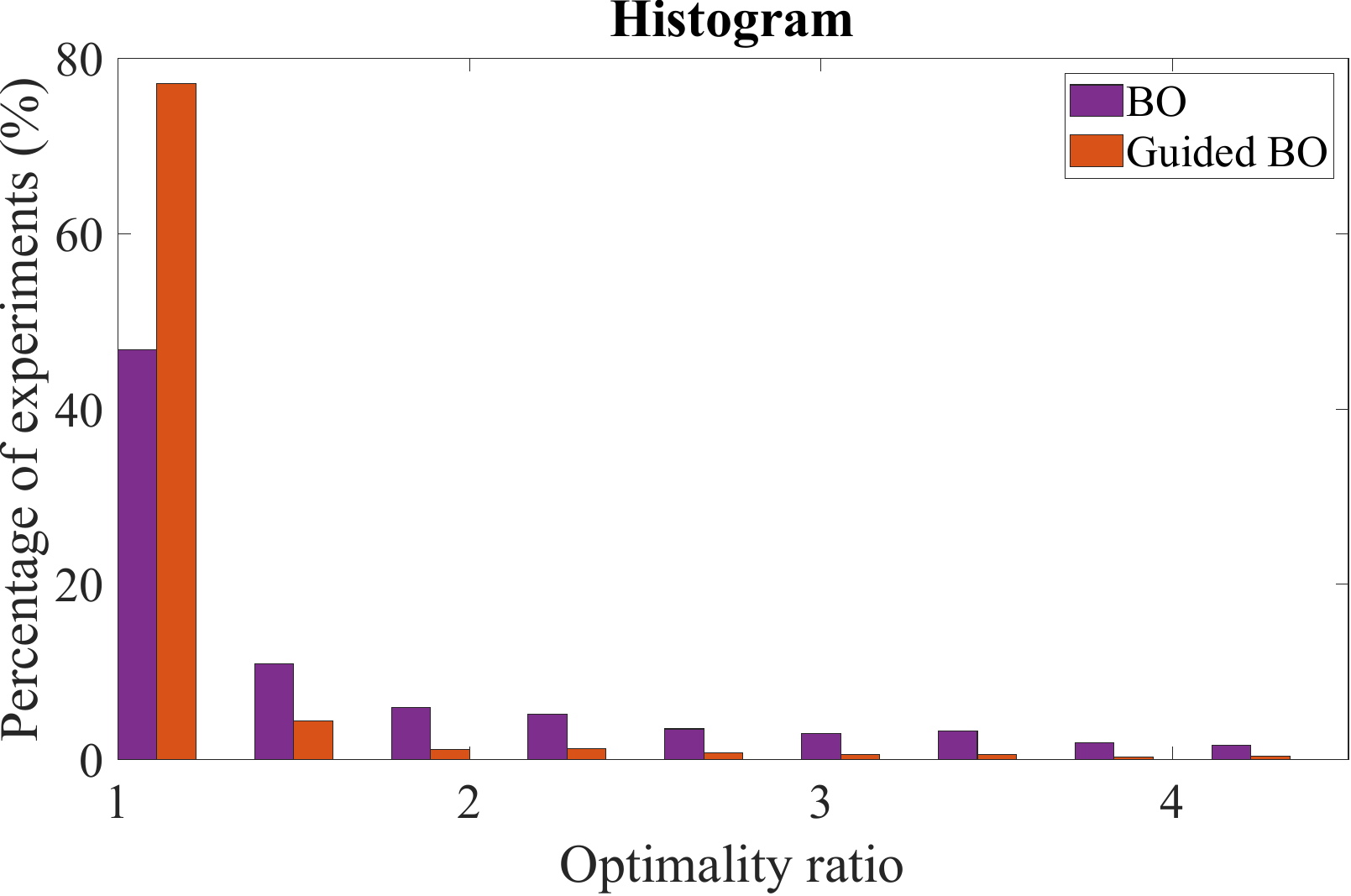}
\caption{Histogram of all measured performance values on the real plant in $100$ batches of $35$ experiments}
\label{fig:numerical_histogram}
\end{figure}

The histogram in Fig. \ref{fig:numerical_histogram} shows the percentage of measurements on the real plant in each optimality ratio range during our Monte Carlo analysis. 
This histogram indicates a higher proportion of the guided BO experiments around the optimum, allowing the optimizer to exploit more around the optimum region.
The BO requires more exploration with higher optimality ratios.

\begin{table}[!h]
\centering
\setlength\extrarowheight{1.1pt}
\caption {\label{table:numerical_gains} Comparison of optimum cost and controller parameters tuned by ground truth grid search, nominal, BO and guided BO methods after $n=21$ iterations on $\mathrm{G}$.} 
\begin{tabular*}{0.45\textwidth}{@{\extracolsep{\fill}}*{6}{c}} \toprule
{Controller} & {$\mathrm{K}_{\text{p}}$} & {$\mathrm{K}_{\text{i}}$} & {$\hat{J}([\mathrm{K}_{\text{p}},\mathrm{K}_{\text{i}}])$} \\ \midrule
    {Ground truth grid search}     & 0.54  &  1.16 & 0.79  \\ 
    {Nominal($\Phi_{\text{m}}=60{°}$)}   & 0.85  &  1.07 & 1.05    \\ 
    {BO}          & 0.45  &	1.18 & 0.84   \\
    {Guided BO}   & 0.53  & 1.16 & 0.79  \\ \bottomrule

\end{tabular*}
\end{table}

Table \ref{table:numerical_gains} specifies the tuned controller parameters and corrosponding cost values.
The nominal controller is the PGM controller introduced in subsection \ref{subsec:nominal_tuning} with phase margin of $60{°}$.
The ground truth optimum controller gains are retrieved by dense grid search in the feasible set on the numerical system.
According to this table, the tuned controller by our guided BO method has lower cost than BO and the nominal controller, and its tuned controller parameters are almost equal to the ground truth optimum gains.

\begin{figure}[!h] 
\centering
\includegraphics[width=3.3in, trim=0cm 0cm 0cm 0cm, clip]{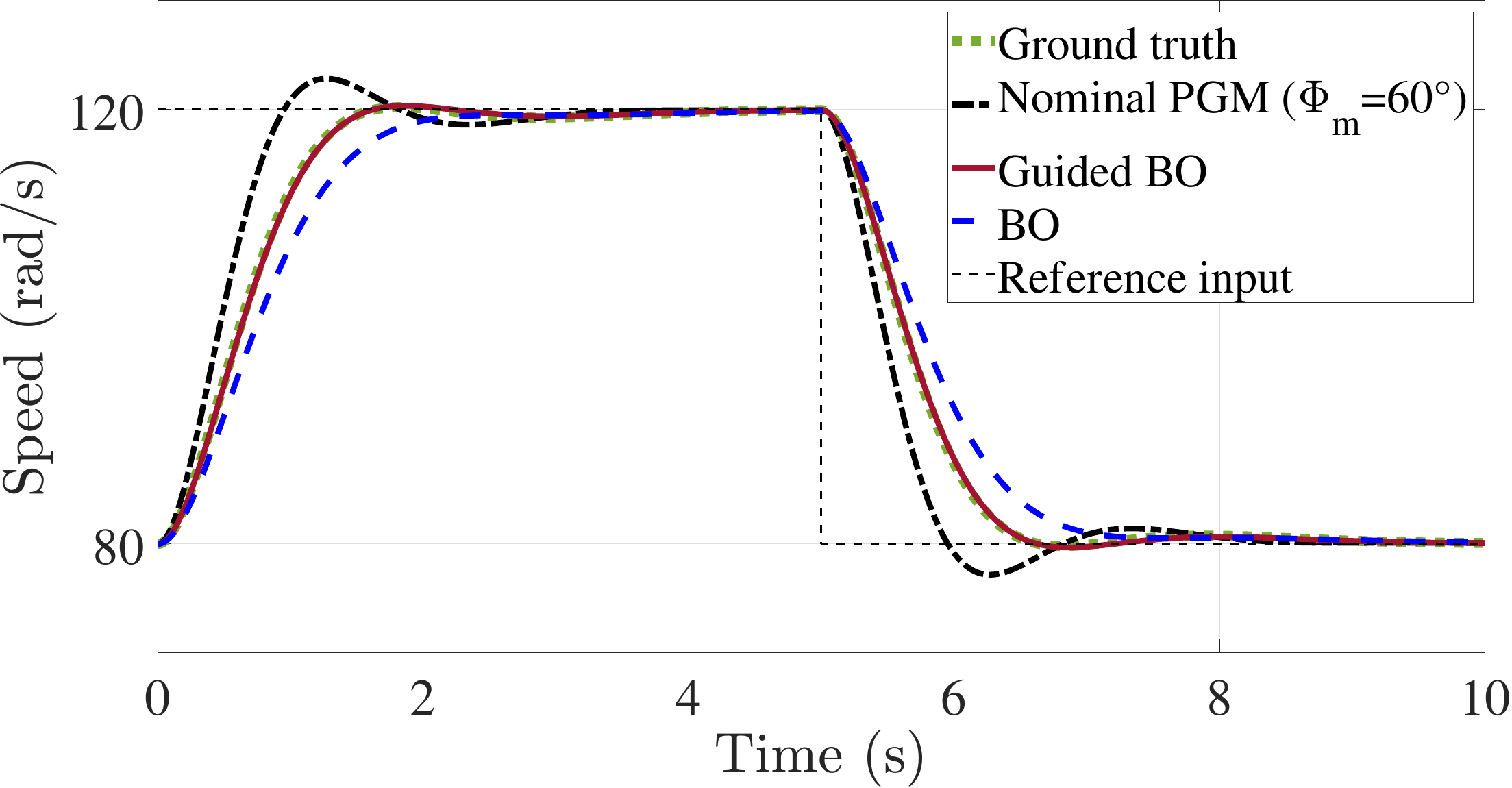}
\caption{Step response of the closed-loop numerical system with the speed controller tuned with different methods}
\label{fig:StepRsps}
\end{figure}

Fig. \ref{fig:StepRsps} shows the step response of the closed-loop system with the guided BO controller on the real plant, corresponding to batch $4$ and experiment $21$ in Fig. \ref{fig:numerical_min}.
While BO step responses have over-damped behavior, the guided BO outperforms the other data-driven and nominal approaches.
However, the nominal has larger overshoot resulting from its large proportional gain.
The guided BO controller performance fits the ground truth performance with a slight overshoot, ITAE, and reasonably fast rise and settling time.

\vspace{-12 pt}
\section{Experimental Results}\label{sec:Exper_Results}
We demonstrate the performance of our proposed guided BO algorithm on two closed-loop real-time systems with different controller and system structures: direct current (DC) rotary motor and linear servomotor systems.
The former system is time-variant, but the latter is time-invariant.
We iteratively tune the speed controller's PI gains for the DC motor while we tune the position controller's PD gains for our servomotor system.
At the end, each method's total controller tuning time is compared. 
\vspace{-10 pt}




\subsection{Rotary Motor System}\label{subsec:rotary_motor_system}
\begin{figure}[h!] 
\centering
\includegraphics[width=3.6in, trim=2.5cm 3.9cm 1.5cm 3cm, clip]{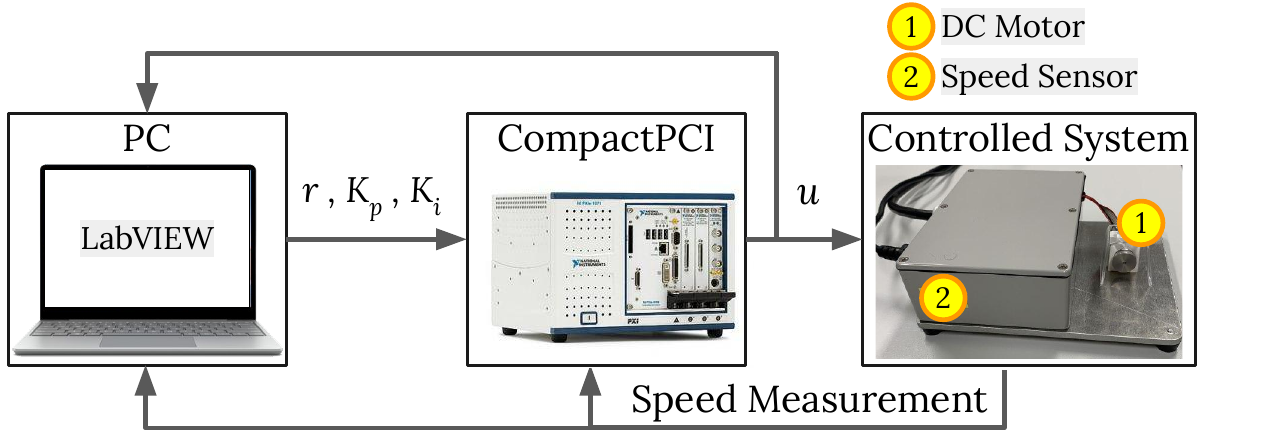}
\caption{\label{fig:DC_motor} Real-time DC rotary motor system structure for measurement and control}
\end{figure}

We have a DC rotary electrical motor setup presented in Fig. \ref{fig:DC_motor}.
This setup allows us to measure the system feedback in real time and modify the closed-loop system behavior by overwriting the controller gains.
The controlled system comprises a DC rotary electrical motor with an angle encoder for speed measurements. 
We use a CompactPCI real-time engine $\text{PXI-7846R}$ from National Instruments Corp to implement the PI speed controller.
Given the reference input signal $r$ and the PI controller gains $\mathrm{K}_{\text{p}}$ and $\mathrm{K}_{\text{i}}$ to the CompactPCI, a digital-to-analog data conversion with pulse-width modulation converts the digital speed control command to the analog voltage signal.
Then, a current converter amplifies the voltage to its proportional current $u$ as an input to the DC motor.
We use LabView software on our PC to interface with the hardware and implement the guided BO algorithm.



We want to tune the PI speed controller gains.
We study the performance of our iterative tuning methods in the presence of unknown effects, such as friction and measurement noise.
We focus on controlling the speed of the motor as it is relevant for its practical use. 
At the beginning of each experiment, we initialize the motor speed to reach a constant encoder angular speed equal to $r_{1}=245$ rad/sec.
This allows us to remove the uncertainty of starting from an arbitrary initial condition on this hardware.
We build the initial data set $\mathcal{D}$ with one gain pair ($N_{0}=1$) and the respective measured objective according to \eqref{eq:J_used}.
With a sampling rate of $20$ Hz, we measure $\tilde{n}=100$ pairs of control commands $u$ and physical plant's output $y$.
In the guided BO algorithm, we use \texttt{tfest} function in MATLAB \cite{MATLAB} with a second-order model to identify the DT of the nonlinear DC motor system based on refined instrumental variable method \cite{IV2015}.
By measuring five times the performance metrics in \eqref{eq:J_used} with a randomnly selected $\theta \in \Theta$, we estimate the additive noise variance $\sigma_{\epsilon}=0.03$.
Table \ref{table:parameter_values} summarizes the guided BO parameter parameter values.

We study the measured performance of $100$ experiment batches.
Each batch starts with a different $\Theta_{\text{init}}$ randomly selected inside the feasible set $\Theta$ in \eqref{eq:PI_gain_bounds}, which is identical for both guided BO and BO methods. We define the optimum cost as the minimum cost measured among all batches of experiments.
The optimality ratio is between the minimum observed cost in each batch and the optimum cost.

\begin{figure}[!h] 
\centering
\includegraphics[width=0.4\textwidth]{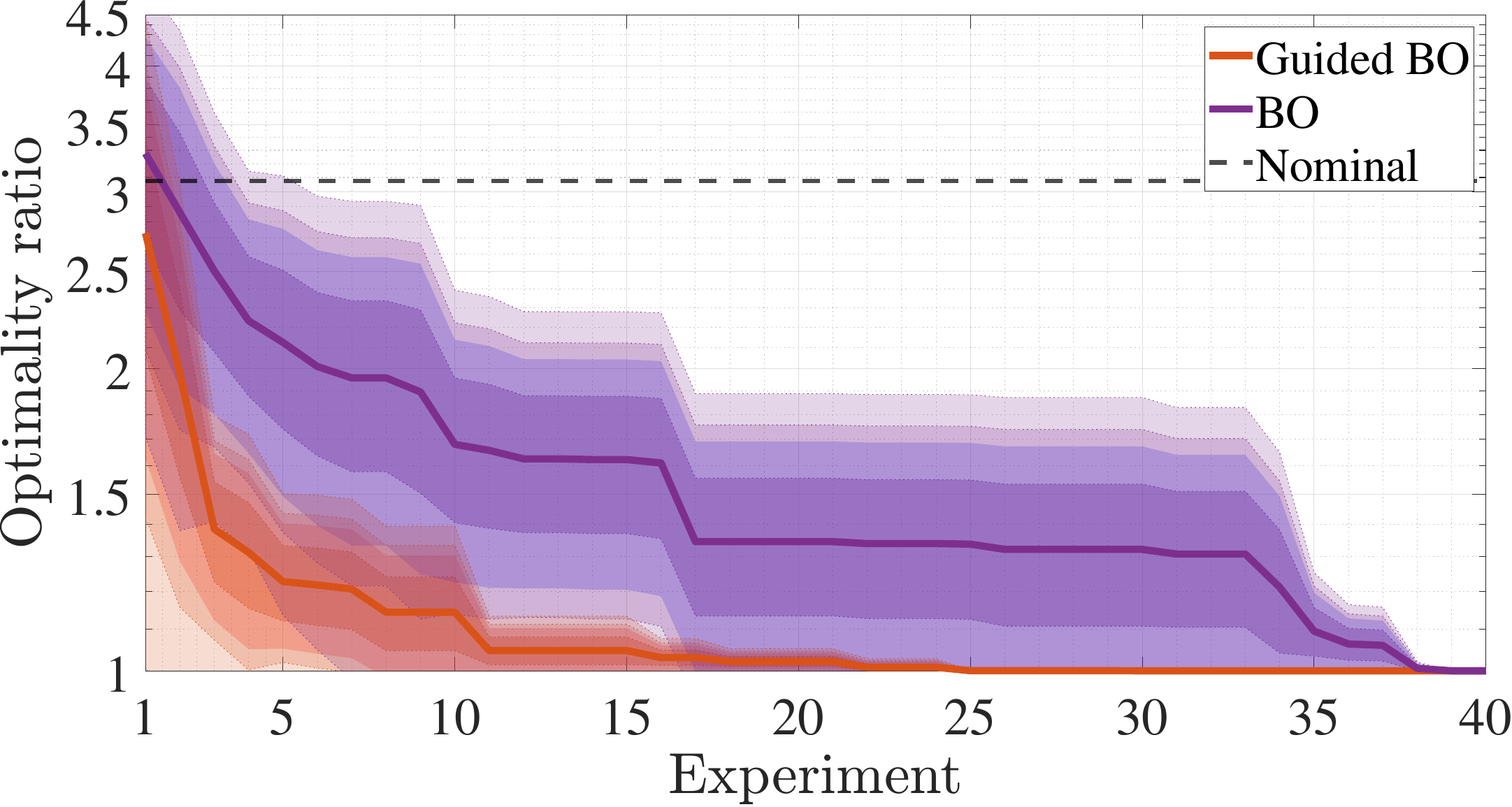}
\caption{Minimum observed optimality ratio up to each experiment on the DC rotary motor system. The thick line is the average over $100$ batches, and the shaded area shows the $99\%$, $95\%$, $90\%$, and $68\%$ confidence intervals.}
\label{fig:Exper_DC_min}
\end{figure}
\vspace{-5pt}
Fig. \ref{fig:Exper_DC_min} demonstrates that guided BO converges on average with fewer experiments on the real DC rotary motor system and has a smaller variance between repetitions.
Let's consider a threshold for the optimality ratio within $5\%$ of the optimal cost.
BO requires, on average, $38$ experiments to converge to the optimality ratio of $1.05$, whereas guided BO requires only $11$ experiments to perform the same, a $71\%$ improvement.
Indeed, the best among the $100$ batches of guided BO accomplishes this in $3$ experiments (compared to $7$ experiments required for BO, a $57\%$ improvement), and the worst in $30$ experiments (compared to $39$ experiments required by BO, a $23\%$ improvement).
After $15$ experiments, all $100$ repetitions of guided BO attain optimality ratios better than $1.1$, compared to $55\%$ of the BO repetitions.
In contrast, at this stage, the worst among the BO repetitions still has an optimality ratio of $2.85$.
The dashed black line in Fig. \ref{fig:Exper_DC_min} depicts the cost value given the first nominal PGM controller designed based on the identified linear model of the system as described in Section \ref{subsec:System_Structure_Identification}.
Both data-driven controller tuning approaches outperform the nominal GPM method in terms of the cost value at the expense of the number of iterations needed.

\begin{figure}[h!] 
\centering 
\includegraphics[width=0.4\textwidth]{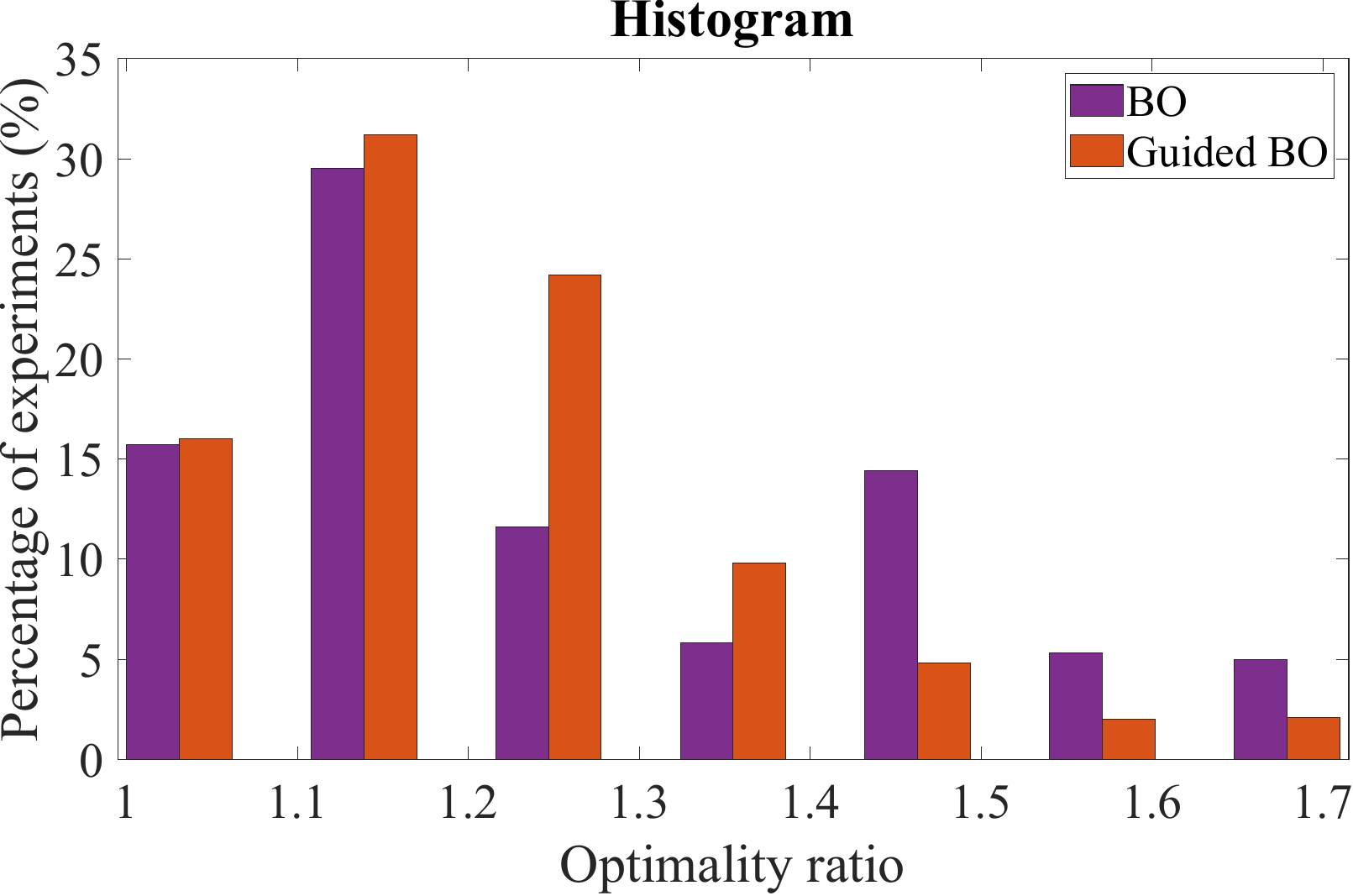}
\caption{Histogram of the experiments percentage vs. the optimality ratio of the measured performance among $100$ batches of $40$ experiments on the rotary motor system.}
\label{fig:Exper_DC_hist}
\end{figure}

From the histogram of all measured performance values on Fig. \ref{fig:Exper_DC_hist}, it can be seen that BO on the actual system requires performance measurements further away from the optimum, whereas the DT carries out the exploration allowing guided BO to test the controller gains on the real system at regions with smaller costs.
Therefore, guided BO exploits more around the optimum cost, whereas BO explores further away from the optimum.
\vspace{-10 pt}
\vspace{-20 pt}
\subsection{Linear Servo Motor System}\label{subsec:linear_servo_motor_system}

\begin{figure}[!h] 
\centering
\includegraphics[width=3.5in, trim=0.3cm 4cm 0.3cm 2cm, clip]{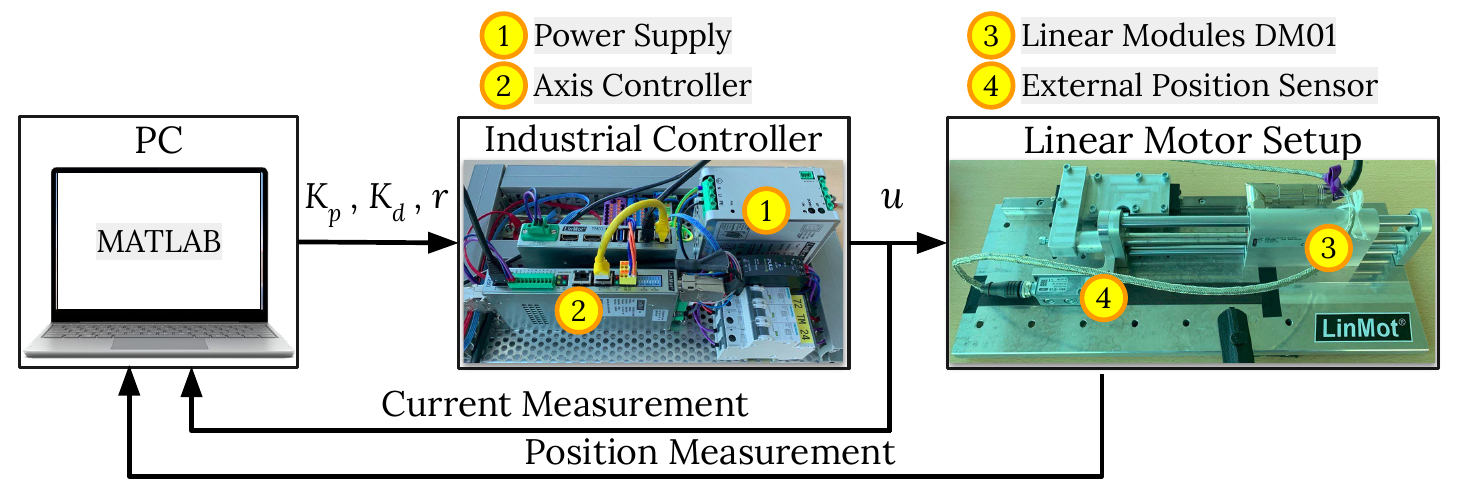}
\caption{\label{fig:LM} LinMot P01 linear servo motor closed-loop control system and data-driven controller tuning framework}
\end{figure}

Our linear motor setup has a DM01 linear module from NTI AG company, displayed in Fig. \ref{fig:LM}.
This module consists of a linear guide with an integrated LinMot P01 linear permanently actuated servo motor.
We integrate a \text{MS01-1/D-SSI} non-contacting external position sensor with our module.
This sensor has $5$ $\mu \text{m}$ absolute resolution and $0.005$ mm repeatability error.
We build a unit feedback closed-loop system with a \text{C1250-IP-XC-1S-C00} NTI AG axis controller and our external sensor to control the output position of the LinMot P01 linear motor with a sampling frequency of $1$ kHz.
This industrial position controller is a PDT1 controller implemented in C and defined as
\begin{equation}\label{eq:LM_Controller}
C(s)=\mathrm{K}_{\text{p}}+\mathrm{K}_{\text{d}}\frac{s}{0.001s+1},
\end{equation}
where $\mathrm{K}_{\text{p}}$ and $\mathrm{K}_{\text{d}}$ are proportional and derivative gains, respectively.
This controller commands the linear motor plant as a current signal.
A saturation mechanism clips the control signal once it exceeds the range $[-7\mathrm{A},7\mathrm{A}]$.
We use OPC Unified Architecture in MATLAB to communicate with the hardware for the input and output measurements and controller parameter tuning.
Given the current and linear motor position measurements, our algorithm sets the reference output $r$ with certain industrial controller gains $\mathrm{K}_{\text{p}}$ and $\mathrm{K}_{\text{d}}$.

We tune $\mathrm{K}_{\text{p}}$ and $\mathrm{K}_{\text{d}}$ gains of the position controller.
To determine the RMSE threshold $\eta_{2}$ in \eqref{eq:DT_rmse1}, we estimate the noise standard deviation by measuring the performance metrics in $5$ experiments with fixed identical controller gains.
We normalize the standard deviation of each metric in $\mathrm{f}=[\zeta, T_{\text{s}}, T_{\text{r}}, e_{\text{ITAE}}]$ according to their mean measured value to retrieve the inverse of the signal-to-noise ratio (1/SNR) equal to $[0.0001, 0.004, 0.002, 0.001]$.
Assuming weighted aggregation-based cost function in \eqref{eq:J_used}, we estimate the additive noise standard deviation as $\sigma_{\epsilon}=0.001$.
We choose the RMSE threshold $\eta_{2}=3\sigma_{\epsilon}=0.003$ to accept DT with sufficient fidelity considering the estimated noise level.



We assume that our linear motor generates linear motion for precise and automatic screw-driving applications.
The speed of the step response is crucial, which can be represented by the rise time performance metric.
One could also expect to avoid overshooting, whereas a slightly longer settling time could be permissible.
We assume the ISO metric thread type M8.1 according to ISO 965 standard for the metric screw thread tolerances \cite{MEStandards}.
This thread type has a Pitch dimension tolerance of $0.112$ mm.
Therefore, considering our position controller's reference step height of $10$ mm, the maximum overshoot must not exceed $\frac{0.112}{10}\times 100 =1.12\%$.

We define the feasible set using a simplified and approximated model of the linear motor system retrieved from the system specifications provided by the LinMot company datasheet \cite{LinMotspecs} where the phase margin is larger than 20 degrees. 
We thus redefine the feasible set for the control parameters as\vspace{-2 pt}
\begin{equation}\label{eq:LM_gain_bounds}
\begin{split}
        \Theta = \Big\{[\mathrm{K}_{\text{p}},\mathrm{K}_{\text{d}}]  \:  \Big|  \: &\mathrm{K}_{\text{p}}\in [\mathrm{K}_{\text{p}_{\min}},\mathrm{K}_{\text{p}_{\max}}],\\
        &\mathrm{K}_{\text{d}}\in[\mathrm{K}_{\text{d}_{\min}}, \mathrm{K}_{\text{d}_{\max}}] \Big\},         
\end{split}
\end{equation}
where the boundaries of each control parameter $\mathrm{K}_{\text{p}_{\min}},\mathrm{K}_{\text{p}_{\max}},\mathrm{K}_{\text{d}_{\min}},\mathrm{K}_{\text{d}_{\max}}$ are equal to $5123.8$, $6136.2$, $40.1$, $51.0$, respectively. 



We approximate the order of magnitude for each performance metric to define our overall performance function. 
We first measure the system's step response given four pairs of control gains in the rectangular feasible set vertices: $(\mathrm{K}_{\text{p}_\text{min}},\mathrm{K}_{\text{d}_\text{min}})$, $(\mathrm{K}_{\text{p}_\text{min}},\mathrm{K}_{\text{d}_\text{max}})$, $(\mathrm{K}_{\text{p}_\text{max}},\mathrm{K}_{\text{d}_\text{min}})$, and $(\mathrm{K}_{\text{p}_\text{max}},\mathrm{K}_{\text{d}_\text{max}})$.
Considering SNR values, we also assume a priori relative importance weight to the metrics suitable for the screw driving application.
Namely, we put $10$ and $5$ percent more weight on the maximum overshoot and rise time metrics.
So the cost function weights $w_{1},w_{2},w_{3},w_{4}$ are equal to $0.10, 0.18, 0.69, 0.04$, respectively, where $\sum_{i=1}^{4}w_{i}=1$.


We compare the performance of our proposed Guided BO with BO in $100$ batches of experiments.
Each batch consists of initial data with $N_{0}$ and $25$ performance measurements on the real linear motor system, whereas Guided BO internally may estimate the cost function with its DT model.
To build the initial data set $\mathcal{D}$ for the batches, we randomly select $100$ pair of gains $\theta = [\mathrm{K}_{\text{p}}, \mathrm{K}_{\text{d}}] \in \Theta$.
We use the same initial data set $\mathcal{D}$ for each batch in both BO and guided BO methods.
In guided BO, each time that we estimate the performance with an updated DT model $\tilde{G}$, we remove previous $(\tilde{\theta},\tilde{\xi})$ data from the training data set $\mathcal{D}$.
We perform an exhaustive grid search in $\Theta$ to calculate the ground truth optimum performance $J(\theta^{*})$ required to calculate the optimality ratios.

\begin{figure}[!h] 
\centering 
\includegraphics[width=0.41\textwidth]{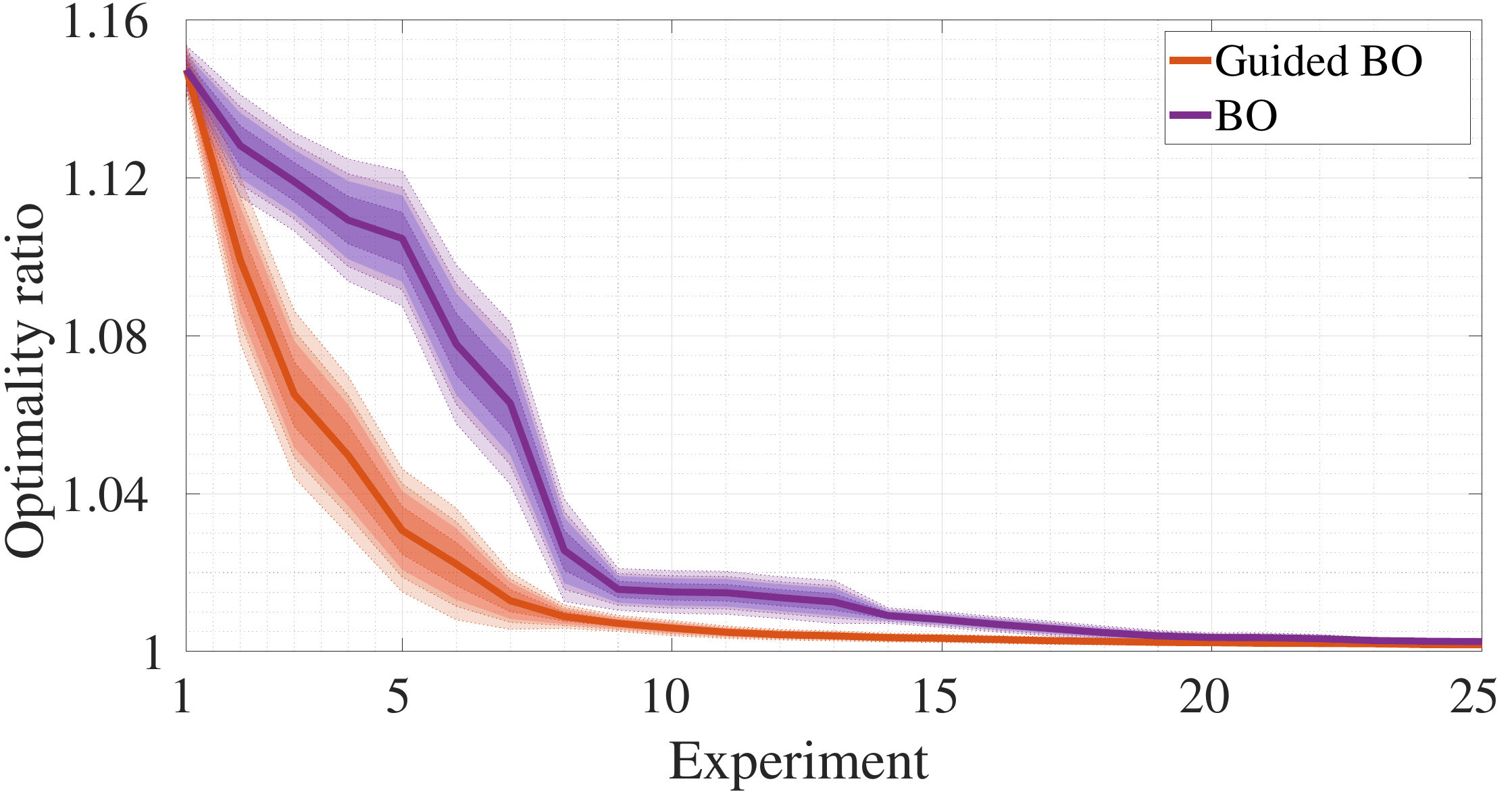}
\caption{Minimum observed optimality ratio on the linear motor system up to each experiment on the linear servo motor system, including $100$ batches of $25$ experiments where only the iterations on the real system are counted. The shaded area shows the $99\%$, $95\%$, $90\%$, and $68\%$ confidence intervals. The thick line is the average of the batches.}
\label{fig:LM_OptRatio_iter}
\end{figure}
\vspace{-5 pt}
Results on Fig. \ref{fig:LM_OptRatio_iter} prove that guided BO, on average, requires fewer experiments on the linear motor system and has a smaller variance between repetitions.
Let's put a threshold on the optimality ratio equal to $1.01$.
BO requires, on average, $18$ experiments on the real plant, excluding the initial data set experiment, to converge to the optimality ratio of $1.01$.; in contrast, guided BO requires only $10$ experiments to perform the same, a $44\%$ improvement.
Indeed, the best among the $100$ batches of guided BO accomplishes this in $2$ experiments (compared to $6$ experiments required for BO, a $66\%$ improvement), and the worst in $17$ experiments, compared to $24$ experiments required by BO, a $29\%$ improvement).
After $10$ experiments, $86\%$ of guided BO iterations attain optimality ratios better than $1.01$, compared to only $14\%$ of the BO repetitions. 


\begin{figure}[!h] 
\centering 
\includegraphics[width=0.4\textwidth]{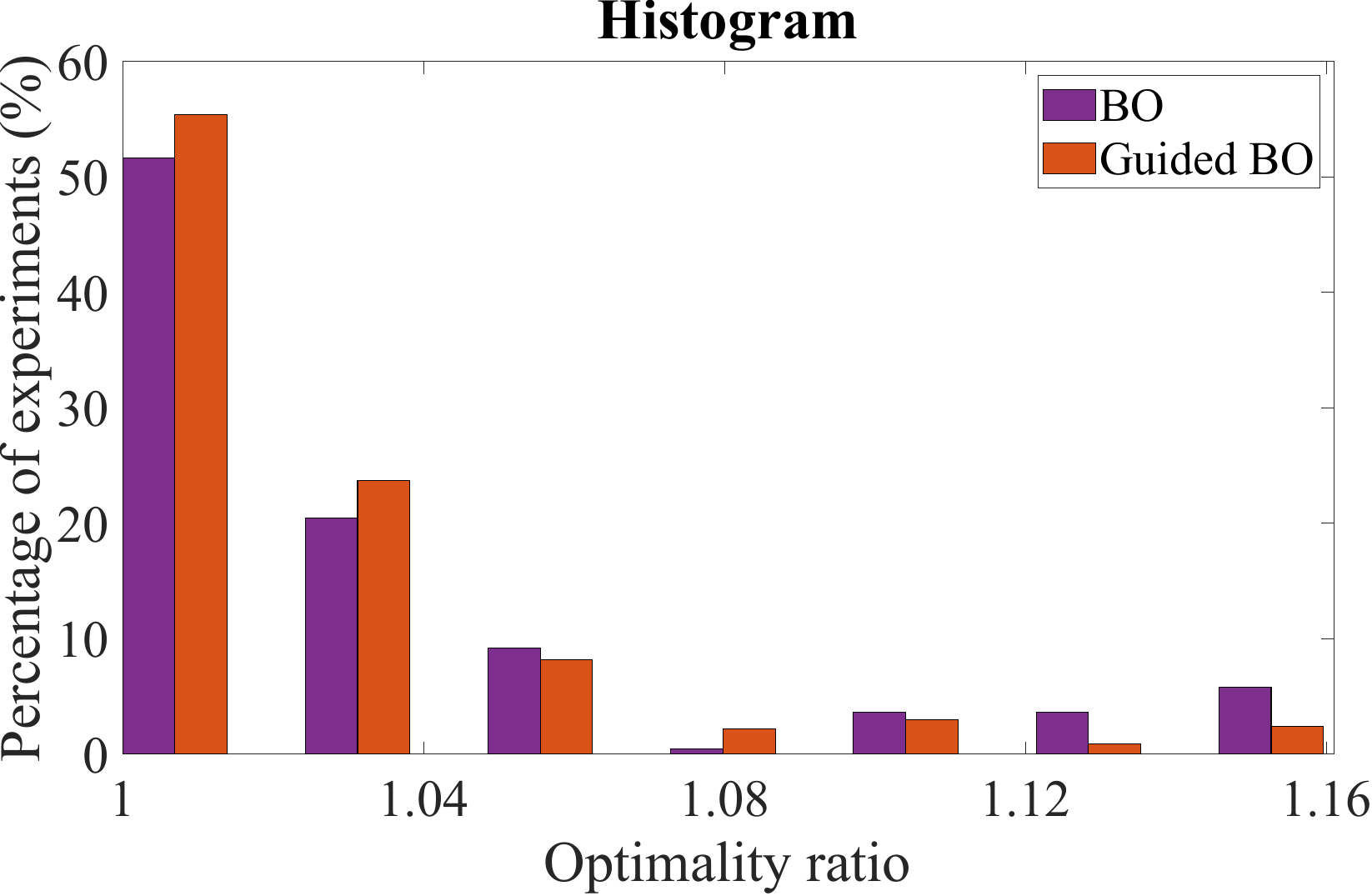}
\caption{Percentage of experiments on the real system vs. the optimality ratio for $100$ batches in a fixed number ($25$) BO iterations.}
\label{fig:LM_hist}
\end{figure}

\vspace{-5 pt}
Furthermore, the confidence interval of the guided BO shrinks faster than BO and both eventually shrink inside the noise boundary.
Looking at the corresponding histogram of all measured performance on the actual system in Fig. \ref{fig:LM_hist}, guided BO does not explore the controller parameters with high-cost values due to the guidance of its DT.
However, BO alone requires further exploration, eventually leading to its slower tuning performance.

\begin{figure}[!h] 
\centering
\includegraphics[width=3.2in, trim=1.7cm  0cm  2.4cm  0cm, clip]{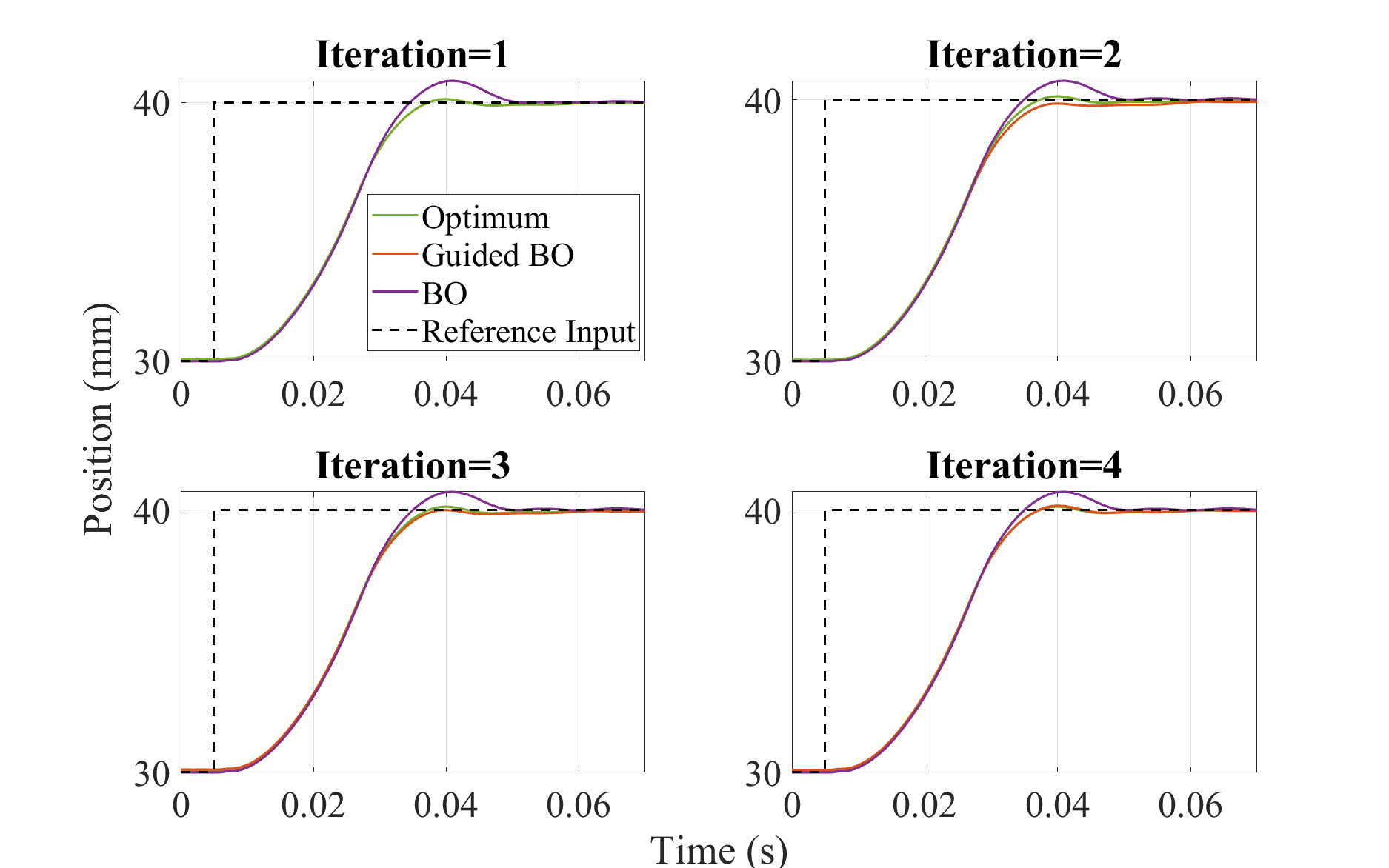}
\caption{Optimum controller performance at different iterations on the real system (linear motor closed-loop step response).}
\label{fig:Exper_LM_step_response}
\end{figure}

We summarize the optimum controller gains of guided BO, standard BO, and the ground truth in Table \ref{table:Exper_LM_opts}.
The guided BO controller's $\mathrm{K}_{\text{d}}$ is higher than the one obtained with BO, in line with the step responses shown in Fig. \ref{fig:Exper_LM_step_response}, causing smaller tracking error rate and a minor overshoot.
The standard BO method has a larger $\mathrm{K}_{\text{p}}$ attempting to respond rapidly immediately after we change to a new reference signal.
The optimum response has a maximum overshoot of $0.06\%$, lower than the required $1.12\%$ by the metric thread standard.
\vspace{-8 pt}
\begin{table}[h!]
\centering
\setlength\extrarowheight{1.1pt}
\caption {\label{table:Exper_LM_opts}Controller parameters for the linear motor setup at iteration $i=13$ of a single batch of experiments in Fig. \ref{fig:LM_OptRatio_iter}.} 
\begin{tabular*}{0.4\textwidth}{@{\extracolsep{\fill}}*{6}{c}} \toprule
{Controller} & {$\mathrm{K}_{\text{p}}$} & {$\mathrm{K}_{\text{d}}$} \\ \midrule
    {Grid search (ground truth)}     & 5341.25 &  49.68    \\ 
    {Guided BO}     & 5138.75 &   49.31   \\
    {BO}     & 6128.75 &  40.81    \\ \bottomrule

\end{tabular*}
\end{table}
\vspace{-1 mm}
Figure \ref{fig:Exper_LM_step_response} visualizes the closed-loop response given the optimum controller gains of guided BO and BO methods after experiment $i=13$ of a batch of experiments.
This figure indicates that the optimum step response tuned by guided BO converges faster than BO to the ground truth optimum response shown with the green line.
We can rationalize the system's behavior based on the aggregated performance metrics. Recalling our cost in \eqref{eq:J_used}, we notice that the area under the curve of guided BO response is less than BO.
While the settling and rise times are similar, we realize that the guided BO reduces the overshoot, resulting in a lower aggregate cost value closer to the optimum response. 
\vspace{-13 pt}

\vspace{-20 pt}

\subsection{Controller Tuning Time}

In each experiment on the physical systems, starting from an initial condition, we measure the step response during $10$ s and $0.30$ s for the rotary and linear motor systems, respectively.
To decrease the noise level, we repeat it twice per experiment for the DC motor setup and take the average estimated value per each performance metric.

Each batch of experiments consists of a single experiment conducted for the offline initial data set and specific experiments per each data-driven tuning, and on average, $121$ experiments on DT in the guided BO method. The DT does not require a sophisticated model that would be time-consuming to calculate, and each BO iteration on the DT requires only $0.09$ s of overdue computation time.
So, the experiments on the physical system demand longer than the DT iterations.

For the PGM nominal controller, one needs to identify the plant model, which takes $7.66$ s for the linear system and $40.35$ s for the rotary system using the Labview-based real-time tool introduced in \cite{KELLER2006177}. The PGM method takes $2$ ms to calculate the controller parameters given the identified LTI model.


\begin{table}[h!]
\centering
\setlength\extrarowheight{1.1pt}
\caption {\label{table:tuning_time} Time required by each controller tuning method} 
\begin{tabular*}{0.4\textwidth}{@{\extracolsep{\fill}}*{5}{c}} \toprule
\multirow{2}{*}{Method} & \multicolumn{2}{c}{Tuning Time [s]}   \\ \cmidrule{2-3}
        &  Rotary system   & Linear system                                               
                                             \\    \midrule
 PGM           &  $40.35$            & $7.67$                \\       
 BO            &   $785.19$          & $16.13$                 \\       
 Guided BO     &  $503.54$           & $14.23$           \\     \bottomrule  

\end{tabular*}
\end{table}

The results of the total tuning time are shown in Table \ref{table:tuning_time}.
Our guided BO is more time-efficient than the BO method because it requires fewer tedious experiments on the physical system and lower computational overdue by DT iterations.

\vspace{-8 pt}
\section{Conclusion}\label{sec:Conclusion}

In this paper, we guided the data-driven Bayesian optimization-based controller with a digital twin according to the uncertainty level of the GP model.
The digital twin is approximated with available data during the system's operation without additional experiments.
As long as the digital twin plant roughly captures the overall behavior of the system far away from the optimum region, it is going to be helpful to guide the Bayesian optimizer.
We showed that the digital twin carries out the exploration duty of the optimizer whereas the exploitation is performed on the real system.
We demonstrated that our guided BO approach considerably improves the controller tuning data efficiency, generalizes across the system and industrial controller structures, and converges faster to optimum values.
Namely, we proved the guided BO's superior performance on a noisy linear servo motor and DC rotary motor real-time hardware.

A prospective extension of our method is to enhance it beyond motor systems to any complex plant with changing behavior.
One can investigate how to develop a constrained version of the guided BO algorithm which can optimally update the feasible set while maintaining the plant's safety properties.
Another future direction is to study estimation methods to obtain the DT plant model by further analyzing excitation signals or various closed-loop identification or regression methods.
Our overall assessment can also be replaced with a multi-objective optimization using Pareto without emphasizing any single performance metric.
Lastly, online validation of the DT model can be integrated with our guided BO method to discard unnecessary information.






\vspace{-5 pt}
\section{Acknowledgement}
We thank Matthias Geissmann for providing the LinMot equipment and useful discussions about its properties.

\vspace{-5 pt}
\bibliographystyle{myIEEEtran}
\bibliography{main}

\vspace{-10 pt}

\section{Biography Section}
\vskip -2.5\baselineskip plus -1fil
\begin{IEEEbiography}[{\includegraphics[width=1in,height=1.25in,clip,keepaspectratio]{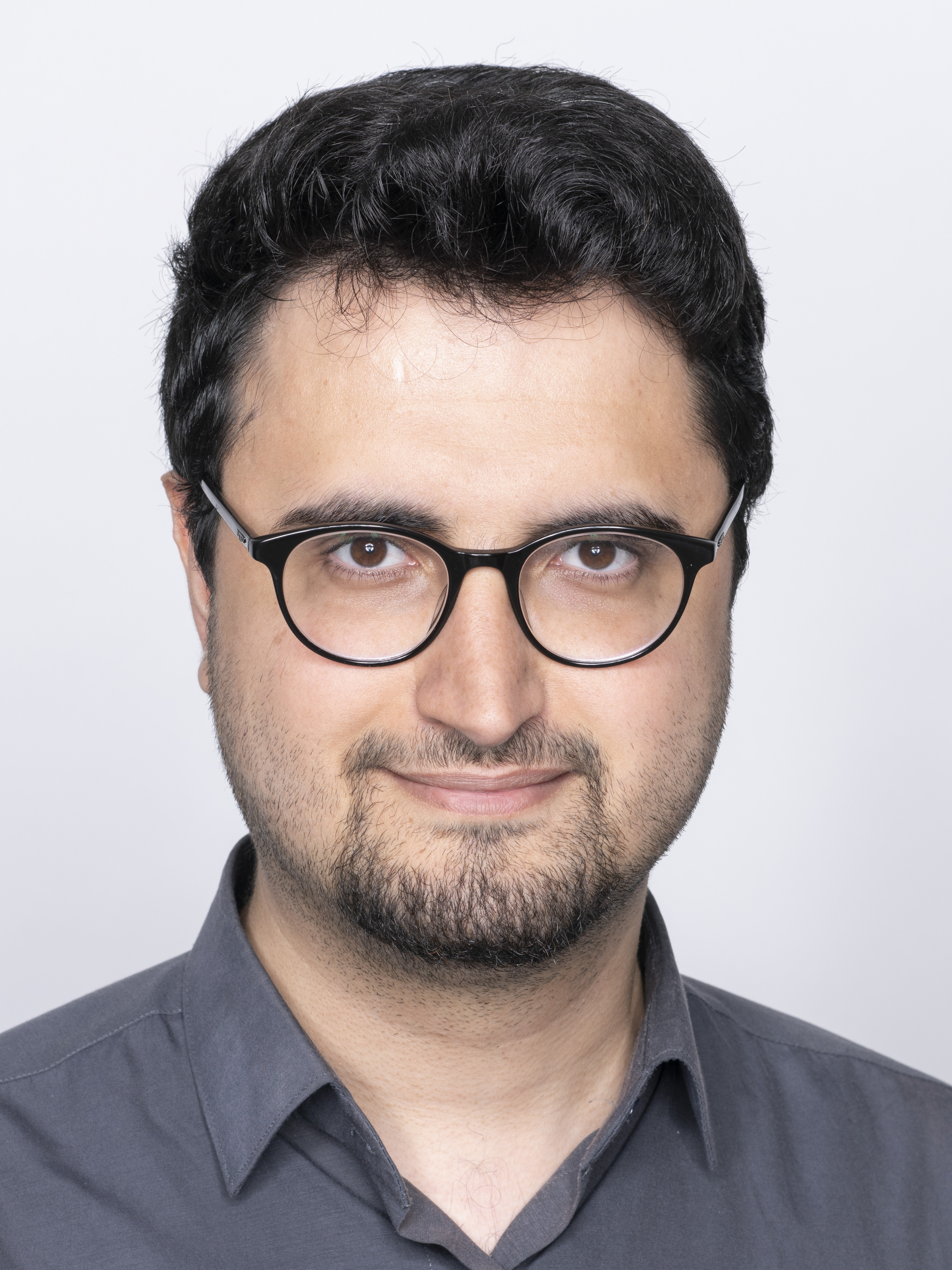}}]{Mahdi Nobar}
is a doctoral candidate at Automatic Control Laboratory, ETH, Zürich, Switzerland. He is also a research assistant at Institut für Automation of the University of Applied Science Northwest Switzerland. Before that, he was a researcher at Robot Learning and Interaction Group, Idiap Research Institute. He obtained his Master's in Mechanical Engineering with a minor in Robotics from EPFL in Lausanne, Switzerland. Mahdi is passionate about solving practical problems where he could provide intelligence to autonomous systems. 
His research interests span data science, optimization, control theories, dynamical systems, and artificial intelligence.
\end{IEEEbiography}
\vskip -4.3\baselineskip plus 5fil

\begin{IEEEbiography}[{\includegraphics[width=1in,height=1.25in,clip,keepaspectratio]{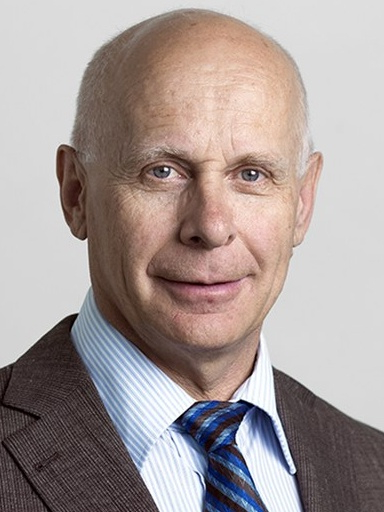}}]{Jürg Keller}
Jürg Keller received his Engineering degree at the ETH Zürich in 1983 and a Ph.D. degree in 1989 at the ETH Zürich. After his postdoctoral work at the ANU, Australian National University, Canberra, he was employed in 1991 at Hoffman-La Roche, Basel, at the Automation of Pharmaceutical Plants department. His main task was to design modular and testable automation software. In 1995, he joined the startup of the University of Applied Sciences in Oensingen/Solothurn, where he had the opportunity to design a bachelor program in automation \& electronics and, in 2002, an automation master program. From 2005 to 2018, he was president of SGA (Schweizerische Gesellschaft für Automatik), the IFAC member organization of Switzerland. Currently, he is a professor of control, machine vision, and machine learning at the FHNW, University of Applied Sciences Northwestern Switzerland. His research interests are applications of modern control methods in industrial environments.
\end{IEEEbiography}
\vskip -2\baselineskip  plus -1fil
\begin{IEEEbiography}[{\includegraphics[width=1in,height=1.25in,clip,keepaspectratio]{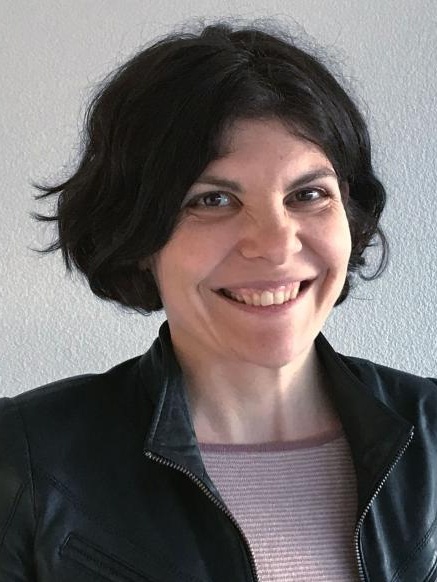}}]{Alisa Rupenyan} (Member, IEEE) received the B.Sc. degree in engineering physics and M.Sc. degree in laser physics from the University of Sofia, Sofia, Bulgaria, in 2004 and 2005, respectively, and the Ph.D. degree from the Department of Physics and Astronomy, Vrije Universiteit Amsterdam, Amsterdam, The Netherlands. She holds the Rieter endowed professorship for Industrial AI at the ZHAW Centre for AI, Zurich University for Applied Sciences, Switzerland. Between 2011 and 2014, she was a Postdoctoral Fellow with ETH Zürich, Zürich, Switzerland, and between 2014 and 2018, she was a Lead Scientist in a robotic start-up. She was a Group Leader in Automation with Inspire AG, the technology transfer unit at ETH Zürich, and a Senior Scientist and PI with the Automatic Control Laboratory, ETH Zurich, between 2018-2023. Her research interests include the intersection between machine learning, control, and optimization for industrial applications and robotics, especially in Bayesian methods applied for optimization and control, as well as learning-based control.
Dr. Rupenyan is an expert for the Swiss Innovation Agency (Innosuisse), a member of several technical committees at IEEE-CSS, IEEE-RAS, and IEEE-IES, and a member of the executive committee at the IFAC Industry Committee.
\end{IEEEbiography}
\vskip -2.5\baselineskip plus -1fil
\begin{IEEEbiography}[{\includegraphics[width=1in,height=1.25in,clip,keepaspectratio]{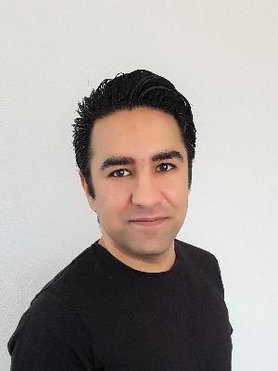}}]{Mohammad Khosravi} is an assistant professor at Delft Center for Systems and Control (DCSC), Delft University of Technology, Netherlands. He received a BSc in electrical engineering and a BSc in mathematical sciences from Sharif University of Technology, Tehran, Iran 2011. He obtained a postgraduate diploma in mathematics from ICTP, Trieste, Italy, in 2012. He was a junior research scientist in the mathematical biology group at Institute for Research in Fundamental Sciences, Iran, from 2012 to 2014. He received his MASc degree in electrical and computer engineering from Concordia University, Montreal, Canada 2016. He obtained his PhD from Swiss Federal Institute of Technology (ETH), Z\"urich, in 2022. He has won several awards, including the ETH Medal, the European Systems \& Control PhD Award, the Outstanding Student Paper Award in CDC 2020, the Outstanding Reviewer Award for IEEE Journal of Control Systems Letters, and the Gold Medal of the National Mathematics Olympiad. His research interests involve data-driven and learning-based methods in modeling, model reduction, optimization, and control of dynamical systems and their applications in thermodynamics, buildings, energy, industry, and power systems.
\end{IEEEbiography}
\vskip -3\baselineskip plus -1fil
\begin{IEEEbiography}[{\includegraphics[width=1in,height=1.25in,clip,keepaspectratio]{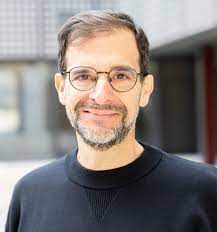}}]{John Lygeros} received a B.Eng. degree in 1990 and an
M.Sc. degree in 1991 from Imperial College, London, U.K., and a Ph.D. in 1996 at the University of California, Berkeley. After research appointments at M.I.T., U.C. Berkeley, and SRI International, he joined the University of Cambridge in 2000 as a lecturer. Between March 2003 and July 2006, he was an Assistant Professor at the Department of Electrical and Computer Engineering, University of Patras, Greece. In July 2006, he joined the Automatic Control Laboratory at ETH Zurich, where he is currently serving as the
Professor for Computation and Control and the Head of the laboratory. His research interests include modeling, analyzing, and controlling large-scale systems with applications to biochemical networks, energy systems, transportation, and industrial processes. John Lygeros is a Fellow of IEEE and a member of IET and the Technical Chamber of Greece. Since 2013, he has been serving as the Vice-President Finances and a Council Member of the International Federation of Automatic Control and, since 2020, as the Director of the National Center of Competence in Research ‘‘Dependable Ubiquitous Automation’’ (NCCR Automation).
\end{IEEEbiography}
\vfill

\end{document}